\documentclass[prd,preprint,superscriptaddress,showpacs,byrevtex]{revtex4}
\usepackage{amsmath}
\usepackage{amssymb}
\usepackage[dvips]{graphicx}
\usepackage{latexsym}

\newcommand{\gtlt}{\stackrel{\scriptstyle >}{<}}
\newcommand{\ltgt}{\stackrel{\scriptstyle <}{>}}

\newcommand{\rd}{\mathrm{d}}
\newcommand{\half}{\frac{1}{2}}
\newcommand{\Tr}[1]{\mathrm{Tr} \{ \, \rho \, #1 \, \} }

\newcommand{\Trquantum}[1]{\mathrm{Tr} [ \, \hat\rho \, #1 \, ] }

\newcommand{\expect}[1]{\langle \, #1 \, \rangle}
\newcommand{\expectz}[1]{\langle \, #1 \, \rangle_{\protect\raisebox{-.7ex}{$\scriptstyle \! 0$}}}
\newcommand{\R}{\mathrm{R}}
\newcommand{\A}{\mathrm{A}}

\newcommand{\C}{\mathcal{C}}
\newcommand{\G}{\mathcal{G}}
\newcommand{\D}{\mathcal{D}}

\newcommand{\calL}{\mathcal{L}}
\newcommand{\calT}{\mathcal{T}}
\newcommand{\calH}{\mathcal{H}}

\begin{document}
%
%
%
\preprint{LAUR 01-3574}
\title{Schwinger-Dyson approach to non-equilibrium classical
       field theory}
\author{Krastan B. Blagoev}
\email{blagoev@physics.bc.edu}
\affiliation{Department of Physics,
       Boston College,
       Chestnut Hill, MA 02167}
\author{Fred Cooper}
\email{fcooper@lanl.gov} \affiliation{Theoretical Division,
   Los Alamos National Laboratory, Los Alamos, NM 87545}
\author{John F. Dawson}
\email{john.dawson@unh.edu}
\affiliation{Department of Physics,
   University of New Hampshire, Durham, NH 03824}
\author{Bogdan Mihaila}
\email{bogdan.mihaila@unh.edu}
\affiliation{Physics Division,
   Argonne National Laboratory, Argonne, IL 60439}

\date{\today}
\begin{abstract}
In this paper we discuss a Schwinger-Dyson [SD] approach for
determining the time evolution of the unequal time correlation
functions of a non-equilibrium classical field theory, where the
classical system is described by an initial density matrix at time
$t=0$.  We focus on $\lambda \phi^4$ field theory in 1+1 space time
dimensions where we can perform exact numerical simulations by
sampling an ensemble of initial conditions specified by the initial
density matrix.  We discuss two approaches.  The first, the bare
vertex approximation [BVA], is based on ignoring vertex corrections to
the SD equations in the auxiliary field formalism relevant for $1/N$
expansions.  The second approximation is a related approximation made
to the SD equations of the original formulation in terms of $\phi$
alone.  We compare these SD approximations as well as a Hartree
approximation with exact numerical simulations.  We find that both
approximations based on the SD equations yield good agreement with
exact numerical simulations and cure the late time oscillation problem
of the Hartree approximation. We also discuss the relationship between
the quantum and classical SD equations.
\end{abstract}
\pacs{11.15.Pg,11.30.Qc,25.75.-q, 3.65.-w}
\maketitle
%
\newpage
%
%
\section{Introduction}
\label{s:introduction}

These past few years, there has been a concerted effort to find
approximation schemes for time evolution problems that go beyond mean
field theory, such as the Hartree\cite{ref:Hartree,ref:GV} or the
leading order in $1/N$ approximations\cite{ref:LOLN}.  Although these
approximations are able to give a reasonable picture of the phase
diagram of quantum field theories (see for example\cite{ref:chodos}),
there are no true hard scatterings in mean field theory, so there is
no mechanism for a system which is out of equilibrium to be driven
back into equilibrium.  Thus various phenomena found in these early
simulations, such as production of disoriented chiral condensates
[DCC's], and distortion of the low momentum spectra away from that of
thermal equilibrium, might very well be modified when hard scatterings
are taken into account.  This is especially true in the
phenomenological $O(4)$ sigma model for the chiral phase transition
where the effective low energy couplings are large.  In Relativistic
Heavy Ion Collider [RHIC] physics applications it is important to
know, given an expanding plasma, how the time scale for equilibration
compares with the time scale for the expansion of the plasma in order
to decide whether effects found in the mean field approximation
persist following the chiral phase transition.

Initial attempts to go beyond mean field theory, such as truncating
the coupled hierarchy of n-point functions at the 4- or 6-point
function level\cite{ref:EQT,ref:abw}, as well as naive use of the
$1/N$ expansion\cite{ref:ctpN,ref:paper1}, have suffered from serious
drawbacks such as negative definite probability and/or secular
behavior.  These problems, although present in naive perturbation
theory, were not present in the mean field approximation because the
mean field approximation can be shown to be equivalent to a
Hamiltonian dynamical system.  Because of these failures, recent
attention has been shifted to resummation methods based on
Schwinger-Dyson [SD] equations resulting from keeping leading terms in
the series for the two particle irreducible [2-PI] generating
functional that occurs in the $\Phi$ derivable approach of
Baym\cite{ref:baym}, or equivalently the formalism of Cornwall,
Jackiw, and Tomboulis [CJT]~\cite{ref:CJT,ref:DM,ref:DJL,ref:VK}. In
this paper we will restrict ourselves to looking at two proposed
resummation methods based on the 2-PI approach. The first approach, is
a direct use of the loop expansion for the 2-PI generating functional
espoused by CJT.  This approach has been used to study thermalization
in the quantum version of the problem discussed here by Berges and
Cox\cite{ref:berges1}.  The second method is to first obtain the exact
SD equations in the auxiliary field formalism\cite{ref:BCG}, and then
to make the assumption that one can ignore vertex corrections.  This
approximation also can be obtained from the 2-PI formalism when we
include propagators for the auxiliary field.  This latter method we
recently applied successfully to the quantum roll problem in 0+1
dimensions\cite{ref:us1}, where we found that the approximation is a
resummation of the next-to-leading order $1/N$ expansion, which cures
the defects of the large-$N$ expansion in next-to-leading order.

In this paper we will apply these approximations to a field theory
problem where exact calculations can be carried out.  Namely we will
consider the time evolution of an initially Gaussian ensemble of
classical fields in 1+1 dimensions.  We will compare the results of
the SD equation approximations with the exact simulation as well as
with the Hartree approximation. To make contact with earlier work in
this area by Aarts, Bonini, and Wetterich\cite{ref:abw}, which went
beyond Hartree approximation by truncating the hierarchy of Green
functions at the four point function level, we will calculate the same
variables used by those authors as well as use their initial
conditions. Since the quantum evolution in the high temperature limit
reduces to this classical field theory problem, by showing that the
bare vertex approximation [BVA] gives reasonable quantitative
agreement for the classical evolution we guarantee that the BVA gives
reasonable results for the \emph{quantum} evolution in the high
temperature regime.

The BVA approximation we espouse here is certainly not new.  It was
first introduced by Kraichnan\cite{ref:Kraichnan} as the ``direct
interaction approximation,'' which he used to study classical and
quantum dynamical systems.  What Kraichnan showed was that this
approximation, like the Hartree, was realizable as a dynamical system
and thus was free from the secularity problem as well as having a
positive definite density matrix.  A method of deriving these
\emph{classical} SD equations was obtained in the classic paper of
Martin, Siggia and Rose [MSR]~\cite{ref:MSR}.  Although quantum field
theorists are now familiar with the Schwinger-Keldysh closed time path
[CTP] formalism for initial value problems\cite{ref:CTP}, the related
formalism of MSR is less familiar, and so we devote an appendix to
deriving the exact SD equations for both \emph{classical} as well as
quantum dynamical systems and discuss the connection between the CTP
and MSR formalisms. A more extensive dicussion of this connection has
just been published\cite{ref:ckr}.

This paper is organized as follows.  In section~\ref{s:theory} we
discuss the classical field theory simulation we are using as a
benchmark for various approximations.  In section~\ref{s:hartree}
we review the Hartree approximation.  In section~\ref{s:BVA}, we
discuss the BVA from various perspectives and compare the quantum
and classical versions.  In section~\ref{s:twoPIeffaction}, we
discuss the 2-PI approximation used by Berges and Cox, and show
that it is a re-expansion of the BVA.  In section~\ref{s:results}
we compare all three approximation schemes with the exact
numerical solutions.  We conclude in section~\ref{s:conclusions}.

In addition, in the appendices we discuss both the CTP and MSR
formalisms for a problem with two interacting fields and discuss the
relationship between the quantum and classical SD equations.
%
%
\section{Classical Scalar Field Theory}
\label{s:theory}

%
%
\subsection{The Lagrangian}
\label{s:lagrangian}

We are interested in studying the classical scalar field theory in
1+1 dimensions described by the Lagrangian density:
\begin{equation}
   \calL
   =
   \half \left  [ \,
      (\partial_t\phi)^2 - (\partial_x \phi)^2 - \mu^2\phi^2 \,
         \right ]
   - \frac{\lambda}{8} \, \phi^4 \>,
   \label{e:lagi}
\end{equation}
with the equation of motion:
\begin{equation}
   \left  [ \,
      \partial_{t}^2
      -
      \partial_{x}^2
      +
      \mu^2 \,
   \right ] \, \phi(x,t)
   +
   \frac{\lambda}{2} \, \phi^3(x,t)
   = 0  \>.
   \label{e:lagieom}
\end{equation}
In momentum space, we define:
\begin{equation}
   \phi(x,t)
   =
   \int_{-\infty}^{+\infty} \frac{\rd k}{2\pi} \, 
   \tilde{\phi}_k(t) \,
   e^{i k x} \>,
   \label{e:fouriertransformdef}
\end{equation}
in which case \eqref{e:lagieom} becomes:
\begin{equation}
   \left [ \, 
   \partial_{t}^2 
   +
   \omega_k^2(t) \,
   \right ] \,
   \tilde{\phi}_k(t)
   +
   \frac{\lambda}{2} 
   \int_{-\infty}^{+\infty}\!\!\! \rd x \, 
   \phi^3(x,t) \,
   e^{-i k x}
   =
   0 \>,
   \label{e:eomfouriertransform}
\end{equation}
where $\omega_k^2 = k^2 + \mu^2$.     
Because we are particularly interested in SD equations which are a
resummation of the large-$N$ expansion, we will also be using a second
form of this Lagrangian:
\begin{equation}
   \calL
   =
   \half \left  [ \,
      (\partial_t \phi)^2 - (\partial_x \phi)^2 - \chi \, \phi^2 \,
         \right ]
   +
   \frac{\chi}{\lambda} \left( \frac{\chi}{2} - \mu^2 \right) \>.
   \label{e:lagii}
\end{equation}
This second Lagrangian leads to the equations of motion:
\begin{equation}
   \left  [ \,
      \partial_{t}^2
      -
      \partial_{x}^2
      +
      \chi(x,t) \,
   \right ] \, \phi(x,t)
   = 0  \>,
   \label{e:eomclassical}
\end{equation}
and the constraint (``gap'') equation for $\chi(x,t)$:
\begin{equation}
   \chi(x,t)
   =
   \mu^2
   +
   \frac{\lambda}{2} \, \phi^2(x,t) \>.
   \label{e:chiclassical}
\end{equation}
The Hamiltonian density is given by:
\begin{equation}
   \calH
   =
   \half \left  [
      (\partial_t\phi)^2 + (\partial_x \phi)^2 + \chi \, \phi^2
         \right ]
   -
   \frac{\chi}{\lambda} \,
   \Bigl (
      \frac{\chi}{2} - \mu^2
   \Bigr ) \>.
   \label{e:hamiltonian}
\end{equation}

%
%
\subsection{Initial ensemble}
\label{s:ensemble}

Solutions of the equations of motion are found for the initial
conditions $\phi(x,0) = \phi(x)$ and $\pi(x,0) = \pi(x)$.
If the system is in thermal equilibrium, the initial
values $\phi(x)$ and $\pi(x)$ are taken from a canonical
ensemble governed by a classical density distribution
$\rho[\phi,\pi]$, given by:
\begin{align}
   \rho[\phi,\pi]
   &=
   Z^{-1}(\beta) \, \exp \{ -\beta H[\phi,\pi] \} \>,
   \notag \\
   Z(\beta)
   &=
   \prod_x \iint
   \rd \phi(x) \, \rd \pi(x) \,
   \exp \{ -\beta H[\phi,\pi] \}  \>.
\end{align}
with $\beta = 1/T$.  The ensemble average of a quantity $A[\phi,\pi]$
is then defined by:
\begin{align}
   \expect{ A[\phi,\pi] }
   &\equiv
   \Tr{ A[\phi,\pi] }
   \notag \\
   &\equiv
   Z^{-1}(\beta)
   \prod_x \iint
   \rd \phi(x) \, \rd \pi(x) \,
   A[\phi,\pi]  \, \exp \{ -\beta H[\phi,\pi] \}  \>,
   \label{e:expectfull}
\end{align}

In order to make direct comparisons with the work of Aarts, Bonini
and Wetterich\cite{ref:EQT}, we choose an ensemble constructed
from unperturbed Hamiltonian $H_0$ rather than $H$ in this paper.
This is \emph{not} a canonical equilibrium distribution for either
the exact solution \emph{or} the Hartree approximation, so it
gives a non-trivial initial value problem for all our
approximations.  In classical 1+1 dimensional field theory, mass
renormalization is finite so that one can use any arbitrary mass
in our initial $H_0$; however, in order to compare our results
with Aarts, \emph{et.~al.}, we take this mass to be the bare
(unperturbed) mass $\mu$.  In dimensions higher than 1+1 one would
need to use a renormalized mass parameter in $H_0$.  So we choose
for our initial density matrix:
\begin{equation}
   \rho_0[\phi_0,\pi_0]
   =
   Z_0^{-1}(\beta_0) \exp \{ -\beta_0 H_0[\phi_0,\pi_0] \} \>,
  \label{e:densityz}
\end{equation}
where $\beta_0 = 1/T_0$ is to be regarded as a parameter.  The
unperturbed Hamiltonian is
\begin{equation}
   H_0[\phi_0,\pi_0]
   =
   \frac{1}{2}
   \int_{-\infty}^{+\infty}\!\!\! \rd x \, 
   \left  [
      \pi_0^2(x) + ( \partial_x \phi_0^{\phantom{2}}(x) )^2
      +
      \mu^2 \, \phi_0^2(x)
   \right ] \>,
   \label{e:H0classical}
\end{equation}
with the equations of motion,
\begin{equation}
   \left  [ \,
      \partial_{t}^2
      -
      \partial_{x}^2
      +
      \mu^2
   \right ] \, \phi_0(x,t)
   = 0 \>,
   \label{e:eom0}
\end{equation}
where $\pi_0(x,t) = \dot\phi_0(x,t)$.  Again, defining Fourier
transforms by:
\begin{equation*}
   \phi(x,t)
   =
   \int_{-\infty}^{+\infty} \frac{\rd k}{2\pi} \, 
   \tilde{\phi}_k(t) \,
   e^{i k x} \>,
   \qquad
   \pi(x,t)
   =
   \int_{-\infty}^{+\infty} \frac{\rd k}{2\pi} \, 
   \tilde{\pi}_k(t) \,
   e^{i k x} \>,
\end{equation*}
solutions for the free particle case are given by:
\begin{align}
   \tilde{\phi}_{0\,k}(t)
   &=
   \phi_k \, \cos(\omega_k t)
   +
   ( \pi_k / \omega_k ) \, \sin(\omega_k t)  \>,
   \notag \\
   &=
   \frac{1}{\sqrt{2\omega_k}} \,
      \Bigl \{ \,
         a_k \, e^{-i\omega_k t} + a_{-k}^{\ast} \, e^{i\omega_k t} \,
      \Bigr \}  \>,
   \notag \\
   \tilde{\pi}_{0\,k}(t)
   &=
   \pi_k \, \cos(\omega_k t)
   -
   \phi_k \omega_k \, \sin(\omega_k t)  \>,
   \notag \\
   &=
   \frac{1}{i} \sqrt{\frac{\omega_k}{2}} \,
      \Bigl \{ \,
         a_k \, e^{-i\omega_k t} - a_{-k}^{\ast} \, e^{i\omega_k t} \,
      \Bigr \}  \>,
\end{align}
where $\omega_k = \sqrt{ k^2 + \mu^2}$.  Note that $\phi(x,0) =
\phi_0(x,0)$ and $\pi(x,0) = \pi_0(x,0)$.  So then
the density matrix becomes:
\begin{equation}
   \rho_0[x_k,y_k]
   =
   \frac{1}{Z_0(\beta_0)} \,
   \exp \biggl \{
      - \beta_0 
      \int_{-\infty}^{+\infty} \frac{\rd k}{2\pi} \, 
      \omega_k \, ( x_k^2 + y_k^2 ) \,
        \biggr \} \>.
   \label{e:densityzi}
\end{equation}
where we have put $a_k = x_k + i y_k$.  So for the Monte Carlo
calculation, we select value $x_k$ and $y_k$ from the Gaussian
distribution in Eq.~\eqref{e:densityzi}, and use these values to
construct starting values for $\phi_0(x,0)$ and $\pi_0(x,0)$.  These
functions are not smooth functions of $x$.  From \eqref{e:densityzi},
we find that:
\begin{gather}
   \expectz{a_k^{\ast} a_{k'}^{\phantom{\ast}}}
   =
   \expectz{a_{k'}^{\phantom{\ast}} a_k^{\ast}}
   =
   2\pi \, \delta(k-k') \, n_k(\beta_0) \>,
   \notag \\
   \expectz{a_k^{\phantom{\ast}} a_{k'}^{\phantom{\ast}}}
   =
   \expectz{a_k^{\ast} a_{k'}^{\ast}}
   =
   0 \>,
   \label{e:expectzaadag}
\end{gather}
and where $n_k(\beta_0) = 1/(\beta_0 \omega_k)$.  Note that
$n_k(\beta_0)$ is the high temperature limit of the classical
Bose-Einstein occupation number distribution.  Then, one show that
$\expectz{\phi(x,0)} = \expectz{\pi(x,0)} = 0$, and that
\begin{alignat}{3}
   \phi_{\text{cl}}^2(0)
   &=
   \frac{1}{L} \int_{0}^{L} \rd x \, \expectz{ \phi^2(x,0) }
   &&=
   \int_{-\Lambda}^{+\Lambda} \frac{\rd k}{2\pi} \, 
   \frac{ n_k(\beta_0) }{ \omega_k }
   &&= 
   \frac{I(\Lambda)}{2 \mu \beta_0} \>,
   \label{e:phi20zfinite} \\
   \pi_{\text{cl}}^2(0)
   &=
   \frac{1}{L} \int_{0}^{L} \rd x \, \expectz{ \pi^2(x,0) }
   &&=
   \int_{-\Lambda}^{+\Lambda} \frac{\rd k}{2\pi} \, 
   n_k(\beta_0) \, \omega_k 
   &&=
   \frac{2 \mu \, J(\Lambda)}{\beta_0} \>,
   \label{e:pi20zfinite}
\end{alignat}
where we have introduced cutoffs $L$ and $\Lambda$ in coordinate and
momentum space integrals.  We have then:
\begin{align}
   I(\Lambda)
   &=
   2\mu
   \int_{-\Lambda}^{\Lambda} \frac{\rd k}{2\pi} \,
   \frac{1}{\omega_k^2}
   =
   \frac{2}{\pi} \
   \tan^{-1} \left ( \frac{\Lambda}{\mu} \right ) 
   \sim
   1 - \frac{2 \mu}{\pi \Lambda} + \dotsb
   \>,
   \label{e:Icontinuum} \\
   J(\Lambda)
   &=
   \frac{1}{2\mu}
   \int_{-\Lambda}^{+\Lambda} \frac{\rd k}{2\pi} 
   =
   \frac{\Lambda}{\pi 2 \mu}  \>.
   \notag 
\end{align}

%
%
\subsection{Correlation and Green functions}
\label{s:correlations}

We define correlation functions and Green functions by ensemble
averages:
\begin{align}
   F(x,x')
   &=
   \expectz{ \phi(x) \, \phi(x') } \>,
   \notag \\
   G_{\R}(x,x')
   &=
   - \sigma(x,x') \, \Theta(t - t') \>,
   \notag \\
   G_{\A}(x,x')
   &=
   + \sigma(x,x') \, \Theta(t' - t) \>,
\end{align}
where $\sigma(x,x')$ is the spectral function, given by the
ensemble average of the classical Poisson bracket of $\phi(x)$ with
$\phi(x')$:
\begin{equation}
   \sigma(x,x')
   =
   \expectz{ \{ \phi(x), \phi(x') \} }  \>.
\end{equation}
[In this section, we use $x \equiv (x,t)$ when needed.]  Since
$\rho_0[\phi_0,\pi_0]$ is Gaussian, the only nontrivial correlation
and Green functions for the free particle case at $t=0$ are the
two-point functions.  These functions are discussed in detail in
Ref.~\cite{ref:parisi}.  Writing $\phi_0(x)$ in a Fourier expansion
and using Eqs.~\eqref{e:expectzaadag}, the unperturbed correlation and
structure functions are given by:
\begin{align}
   F_0(x,x')
   &\equiv
   \expectz{ \phi_0(x) \, \phi_0(x') }
   =
   \int_{-\infty}^{+\infty} \frac{\rd k}{2\pi} \, 
   n_k(\beta_0) \,
   \frac{ \cos[ \omega_k(t - t')]}{\omega_k} \,
   e^{ik(x - x')} \>,
   \notag \\
   \sigma_0(x,x')
   &\equiv
   \expectz{ \{ \phi_0(x), \phi_0(x') \} } 
   =
   - 
   \int_{-\infty}^{+\infty} \frac{\rd k}{2\pi} \, 
   \frac{ \sin[ \omega_k(t - t')]}{\omega_k} \,
   e^{ik(x - x')} \>.
\end{align}

There are similar correlation and Green functions for the $\chi(x)$
field.  Even though this field is a constraint field, we treat it as
if it is dynamic and then take a limit so that the free particle Green
functions, which in this case we call $D_{0\,\R}(x,x')$ and
$D_{0\,\A}(x,x')$, become just delta functions.

%
%
\subsection{Classical solutions on the lattice}
\label{s:classsols}

We introduce a lattice in coordinate and momentum space by putting
$x \rightarrow x_n$ and $k \rightarrow k_m$, where:
\begin{align}
   x_n 
   &= n \, a \>, 
   &
   n &= 0,1, \ldots, N \>,
   \notag \\
   k_m 
   &= 2\pi m / L \>, 
   &
   m &= -N/2, -N/2+1, \ldots, N/2 -1  \>,
   \label{e:latticedefs}
\end{align}
where that $L = N a $ and the momentum cutoff $\Lambda = \pi / a$.  We
write $\phi_n(t) \equiv \phi(x_n,t)$ and $\tilde{\phi}_{k_m}(t) \equiv
\tilde{\phi}_m(t)$.  Fourier transform relations become:
\begin{equation*}
   \phi_n(t)
   =
   \frac{1}{L} \sum_{m=-N/2}^{N/2-1} 
   \tilde{\phi}_m(t) \, e^{ 2\pi i n \, m / N}  
   \>, \qquad
   \tilde{\phi}_m(t)
   =
   a \sum_{n=0}^{N-1} 
   \phi_n(t) \, e^{ -2\pi i n \, m / N} \>.
\end{equation*}

There are two strategies we can use to step out solutions.  In
coordinate space, we can introduce a time grid, $t_s = s a_0$, $s =
0,1,\ldots$, and replace the differential operators for space and time
by second order difference formulas:
\begin{align}
   \partial^2_x \phi(x,t)
   &\rightarrow
   \Bigl [ \,
      \phi_{n+1,s} - 2 \, \phi_{n,s} + \phi_{n-1,s} \,
   \Bigr ] \, / \, a^2   \>, 
   \notag \\
   \partial^2_t \phi(x,t)
   &\rightarrow
   \Bigl [ \,
      \phi_{m,s+1} - 2 \, \phi_{m,s} + \phi_{m,s-1} \,
   \Bigr ] \, / \, a_0^2   \>,
\end{align}
So that Eq.~\eqref{e:lagieom} becomes:
\begin{align}
   \phi_{m,s+1} 
   =
   2 \, \phi_{m,s} - \phi_{m,s-1}
   -
   a_0^2 \, \biggl \{ \,
        \mu^2 \, \phi_{m,s} + \frac{\lambda}{2} \, \phi^3_{m,s} \,
            \biggr \}
   \notag \\ \qquad
   +
   \biggl ( \, \frac{a_0}{a} \biggr )^{\!\!2} \,
   \{ \,
      \phi_{n+1,s} - 2 \, \phi_{n,s} + \phi_{n-1,s} \,
   \} \>.
   \label{e:phicoordlattice}
\end{align}
This method is called a ``staggered leapfrog'' approximation in the
literature\cite{ref:NumRec}, and is believed to be stable.  The
advantage of this coordinate space method on the lattice is that the
interaction term, $\phi^3(x,t)$, is easy to calculate.  We construct
an average of all Monte Carlo runs $M$ for all values of $x_n$ and
compute:
\begin{equation}
   \phi_{\text{cl}}^2(t_s)
   =
   \frac{1}{M N} \sum_{i=1}^{M} \sum_{n=0}^{N-1} \phi_{n,s}^2 
   \sim
   \frac{1}{L} \int_{0}^{L} \rd x \, \expectz{ \phi^2(x,t_s) } \>.
\end{equation}

The use of the second order difference formula means that the
dispersion relation is now given by:
\begin{equation}
   \omega_k^2
   \rightarrow
   \hat{\omega}_m^2 
   = 
   \hat{k}_m^2 + \mu^2  \>, \qquad
   \hat{k}_m 
   = 
   \frac{2}{a} \sin \left ( \frac{\pi m}{N} \right ) \>.
   \label{e:dispspace}
\end{equation}
So at $t=0$, $\expectz{ \phi^2(x,0) }$ is still given by
Eq.~\eqref{e:phi20zfinite} but $I(\Lambda)$ is now given by:
\begin{equation}
   I(\Lambda)
   =
   \frac{2\mu}{L}
   \sum_{m=-N/2}^{N/2-1}
   \frac{1}{\hat{\omega}_m^2}
   =
   \frac{1}{\sqrt{1 + ( \pi \mu / 2 \Lambda )^2 } } 
   \sim
   1 
   - 
   \frac{1}{2} \biggl ( \frac{\pi \mu}{2 \Lambda} \biggr )^{\!\!2} 
   +
   \dotsb \>.
   \label{e:Icoordspace}
\end{equation}
We call this method ``coordinate-space Monte Carlo.''

The second strategy, is to solve Eq.~\eqref{e:eomfouriertransform} in
momentum space, and use a fast Fourier transform to compute the
convolution integral.  Any time integrator can be used for the time
step, but we employ the same second order finite difference formula as
above.  Here, there is no use of the
second order difference formula, so the dispersion relation is
given by:
\begin{equation}
   \omega_k^2
   \rightarrow
   \omega_m^2 
   = 
   k_m^2 + \mu^2  \>, \qquad
   k_m 
   = 
   \frac{2\pi m}{L} \>.
   \label{e:dispmomentum}
\end{equation}
$\expectz{ \phi^2(x,0) }$ is again given by Eq.~\eqref{e:phi20zfinite}
but $I(\Lambda)$ is now given by:
\begin{align}
   I(\Lambda)
   &=
   \frac{2\mu}{L}
   \sum_{m=-N/2}^{N/2-1}
   \frac{1}{\omega_m^2}
   =
   \biggl ( \frac{\pi \mu N}{2 \Lambda} \biggr )
   \sum_{m=-N/2}^{N/2-1}
   \frac{1}{(\pi m)^2 + ( \pi \mu N / 2 \Lambda )^2 }
   \notag \\
   &\sim
   \frac{2}{\pi} \
   \tan^{-1} \left ( \frac{\Lambda}{\mu} \right ) \>,
   \label{e:Imomentumspace}
\end{align}
for sufficiently large $N$.

We call this method ``momentum-space Monte Carlo.''  Since our BVA
codes use a global-in-time Chebyshev expansion method in momentum
space, with a dispersion relation given by Eq.~\eqref{e:dispmomentum},
we can compare directly the results of the BVA code to the classical
momentum-space Monte Carlo calculations.

%
%
\section{Hartree Approximation}
\label{s:hartree}

The Hartree approximation is a simple truncation scheme which is
obtained by setting all correlation functions beyond the second
one to zero.  Since it has a simple interpretation in terms of a
time-dependent variational approximation where the density matrix
is assumed to be Gaussian, it automatically corresponds to a
Hamiltonian dynamical system so it is free from unphysical
behavior.  Here we discuss the classical version of this
approximation.  We are interested in ensemble averages of the
classical evolution.  The equation for the expectation value
$\phi_c(x) = \expectz{\phi(x)}$ is:
\begin{equation}
   \left  [ \,
      \partial_{t}^2
      -
      \partial_{x}^2
      +
      \mu^2
   \right ] \, \phi_c(x)
   +
   \frac{\lambda}{2} \, \expectz{\phi^3(x)}
   = 0 \>.
\end{equation}
[Here $x \equiv (x,t)$.]  Since the Hartree approximation is
equivalent to an initial Gaussian ensemble staying Gaussian, the
expectation values obey the factorization condition
\begin{equation*}
   \expectz{\phi^3(x)}
   =
   \phi_c^3(x) + 3 \, G(x,x) \, \phi_c(x)
\end{equation*}
where
\begin{equation*}
   G(x,x')
   =
   \expectz{\phi(x) \phi(x')}
   -
   \expectz{\phi(x)} \, \expectz{\phi(x')} \>.
\end{equation*}
This equation must be supplemented by the equation for the
expectation value of the fluctuation, which under the Gaussian
hypothesis obeys:
\begin{equation}
   \biggl \{ \,
      \partial_{t}^2
      -
      \partial_{x}^2
      +
      \mu^2
      +
      \frac{3 \lambda}{2} \,
      \Bigl \{ \,
         \phi_c^2(x) + G(x,x)
      \Bigr \} 
   \biggr \} \, G(x,x')
   = 0 \>.
\end{equation}
We will determine the 2-point function by assuming that we can write a
mode decomposition for $\phi(x,t)$ of the form:
\begin{equation}
   \phi(x)
   =
   \phi_c(x)
   +
   \int \frac{\rd k}{2\pi} \,
   [ \, a_k f_k(x) + a^\ast_k f_k^{\ast}(x) \, ]
\end{equation}
where the complete orthonormal set of mode functions $f_k(x)$ obey
\begin{equation}
   \biggl \{ \,
      \partial_{t}^2
      -
      \partial_{x}^2
      +
      \mu^2
      +
      \frac{3 \lambda}{2} \,
      \Bigl \{ \,
         \phi_c^2(x) + G(x,x)
      \Bigr \} 
   \biggr \} \, f_k(x)
   = 0 \>.
\end{equation}
The ensemble average of all the bilinears of $a_k^{\phantom{\ast}}$
and $a_k^{\ast}$ are determined by the initial ensemble average.  For
our classical evolution we have
\begin{equation}
   \expectz{a_k^{\ast} a_{k'}^{\phantom{\ast}}}
   =
   \expectz{a_{k'}^{\phantom{\ast}} a_k^{\ast}}
   =
   2\pi \, n_k(\beta_0) \, \delta(k-k') \>.
\end{equation}
where, again, $n_k(\beta_0) = 1/(\beta_0 \omega_k)$, with $\omega_k =
\sqrt{k^2 + \mu^2}$.

We also need to specify the initial values of $f_k(x)$ and
$\dot{f}_k(x)$ to complete the calculation.  These are determined from
the initial averages given in Eq.~\eqref{e:expectzaadag} and
$\expectz{\phi\phi}$, $\expectz{\pi\pi}$, and $\expectz{\phi\pi}$ at
$t=0$.  We find:
\begin{equation}
   f_k(0)
   =
   \frac{e^{ikx}}{\sqrt{2 \omega_k}} \>, \qquad
   \dot{f}_k(0)
   = - i \omega_k \, f_k(0) \>.
   \label{e:initialfdotf}
\end{equation}

To summarize, for the translationally invariant case when
$\phi_c(x,t)
$ is independent of $x$, the equations
that need to be solved simultaneously are:
\begin{gather}
   [ \, \partial_t^2 + \chi_1(t) \, ] \, \phi_c(t) = 0 \>,
   \qquad
   [ \, \partial_t^2 + k^2 + \chi_2(t) \, ] \, f_k(t)
   = 0 \>,
   \notag \\
   \chi_1(t) = \mu^2 + \frac{\lambda}{2} \phi_c^2(t)
   +
   \frac{3 \lambda}{2} \,
   \int_{-\infty}^{+\infty} \frac{\rd k}{2\pi} \,
   2 n_k(\beta_0) \, | f_k(t) |^2 \>,
   \notag \\
   \chi_2(t) = \mu^2 + \frac{3 \lambda}{2} \phi_c^2(t)
   +
   \frac{3 \lambda}{2} \,
   \int_{-\infty}^{+\infty} \frac{\rd k}{2\pi} \,
   2 n_k(\beta_0) \, | f_k(t) |^2 \>.
\end{gather}
The ensemble average of the fields is given by:
\begin{equation}
   \expectz{\phi^2(t)}
   =
   \phi_c^2(t) +
   \int_{-\infty}^{+\infty} \frac{\rd k}{2\pi} \,
   2 n_k(\beta_0) \, | f_k(t) |^2 \>.
\end{equation}
We remark here that, with the replacement $2 n_k + 1 \rightarrow 2
n_k$, the evolution equations for the Hartree approximation in quantum
mechanics is the same as the classical ones.  In addition, in the
quantum case, in thermal equilibrium, $n_k$ is given by a
Bose-Einstein occupation distribution.  Quantum mechanics also affects
the initial conditions of the mode functions, whose Wronskian, in the
quantum case, must yield $i\hbar$.

%
%

For completeness we mention here that in the quantum case, the
effective action for the Hartree approximation is
\begin{equation}
   S_{\text{eff}}
   =
   \int \rd x \, \calL_{\text{cl}}(\phi,\chi_2)
   + \frac{i}{2} \, \mathrm{Tr} [ \, \ln (\Box + \chi_2) \, ] \>,
\end{equation}
where $\calL_{\text{cl}}(\phi,\chi_2)$ is the classical action written
in terms of the auxiliary field $\chi_2$, given by:
\begin{align*}
   \calL_{\text{cl}}(\phi,\chi_2)
   &=
   - \frac{1}{2} \, \phi(x) \, [ \, \Box + \chi_2(x) \, ] \, \phi(x)
   +
   \frac{\lambda}{4} \, \phi^4(x)
   +
   \frac{1}{3\lambda} \,
   \left [ \, \frac{\chi_2^2(x)}{2} - \mu^2 \, \chi_2(x) \, \right ]
   \>,
   \\
   \chi_2(x)
   &=
   \mu^2 + \frac{3\lambda}{2} \, \phi^2(x) \>,
\end{align*}
Here the $3$ in the Hartree approximation occurring in front of
$\lambda/2$ would become $1+\frac{2}{N}$ when there are $N$
fields, showing that the Hartree reduces to the large-$N$
approximation at large $N$.

We notice that unlike the case of the large-$N$ expansion, the
classical Lagrangian still has a quartic term which has to be
treated as an \emph{external} classical field when doing the
remaining quadratic path integral over $\phi$ to obtain the above
determinant term in the effective action.  Thus, when we take
expectation values of the Lagrangian to obtain the energy momentum
tensor, the quartic term only gets a classical contribution and
does not give fluctuation contributions.
In taking the expectation value in the Hartree approximation one
treats $\chi$ as well as the quartic term in $\phi$ classically.

In terms of the mode functions, the energy density is given by:
\begin{align}
   \varepsilon
   &=
   \frac{1}{2} \,
   \left  [ \,
      (\partial_t \phi^c)^2
      +
      (\partial_x \phi^c)^2
      +
      \chi_2 \, \phi_c^2 \,
   \right ]
   -
   \frac{1}{3\lambda} \,
   \left [ \, \frac{\chi_2^2(x)}{2} - \mu^2 \, \chi_2(x) \, \right ]
   -
   \frac{\lambda}{4} \, \phi_c^4
   \notag \\ & \qquad
   + \frac{1}{2}
   \int_{-\infty}^{+\infty} \frac{\rd k}{2\pi} \,
   2 n_k(\beta_0) \, \left  [ \,
      | f_k(t) |^2
      +
      (k^2 +  \chi_2) \, | f_k |^2 \,
             \right ]
\end{align}
Because classically $n_k \omega_k = T_0$ represents the equipartition
of energy, the classical energy density suffers a linear divergence.
In the quantum case, the Bose-Einstein distribution cures this, and
instead one obtains a quadratic divergence from the zero point quantum
fluctuations, which must be renormalized by a constant counterterm,
corresponding to a cosmological constant.

%
%
\section{Bare Vertex Approximation}
\label{s:BVA}

The Hartree approximation truncates what would be otherwise an
infinite set of coupled equations for the expectation values of
products of the classical field by assuming that all fluctuations
higher than the second one are exactly zero.  In quantum mechanical
applications, this approximation has many nice features such as being
derivable from a variational approximation and having a well defined
positive definite density function at all times. However, it suffers
from not containing the hard scatterings usually associated with
thermalization.  It is thus important to find approximations which
include scattering and which are free from the problems of secularity
and positivity as we discussed in the introduction.

From the generating functional of all the correlation functions one
finds that the Green functions for both quantum and classical field
theory obey a set of coupled integral equations relating the exact two
point functions to themselves and higher order one-particle
irreducible vertex functions.  In quantum field theory, the easiest
way to derive these is by using a generating functional for the matrix
Green functions of the CTP formalism.  The classical theory has
similar structure to the Green functions found when we tridiagonalize
the matrix Green functions and write SD equations for the retarded
Green functions and Wightman functions.  The classical limit ($\hbar
\rightarrow 0$) of the BVA is obtained by using the classical value
for the free retarded Green functions and Wightman function and also
only keeping those graphs which are leading at high temperature and
small coupling as discussed by Aarts and Smit\cite{AaSm97} and by
Buchm\"uller and Jakov\'ac\cite{ref:BuJa97,ref:BuJa98}.  For the classical
theory, one can define free propagators which replace the quantum ones
by replacing commutators by Poisson brackets, as discussed by
Parisi\cite{ref:parisi}.  In appendix~\ref{s:MSRtheory}, we show how
to make precise the connection between the exact quantum and classical
Schwinger-Dyson equations, by comparing the CTP formalism with the MSR
formalism.  We find that certain bare vertices are missing in the
classical case.

The BVA is an approximation in which the exact two point functions are
used in the equations for $\expectz{\phi(x)}$ and $\expectz{\phi(x)
\phi(x')}$, but the exact $\expectz{\phi(x) \phi(x') \chi(x'')}$
vertex function is replaced by the bare one.  This approximation was
discussed from several points of view in our previous
paper\cite{ref:us1}, where we showed that the BVA is a particular
resummation of the $1/N$ expansion. For the classical problem we treat
here, the exact SD equations are obtained from the MSR
formalism\cite{ref:MSR}. We sketch the details of this in
appendix~\ref{s:MSRtheory}. What is important about the BVA is that
this approximation is energy conserving, is nonsecular and, as
Kraichnan has shown\cite{ref:Kraichnan}, the effective Hamiltonian has
positive spectra.

In condensed matter physics the BVA is used in the electron-phonon
problem for the description of the normal state of a Fermi liquid.  In
this case the first vertex correction to the self-energy is small
compared to the bare vertex and are of the order of $m_e/M_{ion}$
where $m_e$ is the electron mass and $M_{ion}$ is the ion mass. This
result is known as the Migdal theorem\cite{ref:Migdal58}. Close to a
superconducting instability the theorem is not valid and a partial
resummation of the vertex corrections is
necessary\cite{ref:Schrieffer64}.  The physical reason for the
validity of Migdal's theorem is the adiabatic motion of the electronic
degrees of freedom compared to the ionic degrees of freedom.  In
quantum field theory it is not yet clear what the domain of validity
of the BVA will be.  Here we are concerned with its validity in the
high temperature limit.

%
%
\subsection{The quantum Schwinger-Dyson equations in the BVA}
\label{s:SDinBVA}

The exact SD equations for $\lambda \phi^4$ field theory,
rewritten in terms of an auxiliary field, were first derived by
Bender, Cooper and Guralnik\cite{ref:BCG} and discussed in detail
in the CTP formalism in Ref.~\cite{ref:us1}. We will also derive
the SD equation for both the classical and quantum cases in the
appendix.  The CTP formalism and notation is reviewed in
appendix~\ref{s:CTPtheory}.  For the case where $\phi(x) =
\expectz{\hat\phi(x)} = 0$, one has that
\begin{align}
   D(x,x')
   &=
   D_0(x,x')
   -
   \int_{\C} \rd^2 x_1 \!\! \int_{\C} \rd^2 x_2 \,
   D_0(x,x_1) \, \Pi(x_1,x_2) \, D(x_2,x') \>,
   \label{e:SDD}
   \\
   G(x,x')
   &=
   G_0(x,x')
   -
   \int_{\C} \rd^2 x_1 \!\! \int_{\C} \rd^2 x_2 \,
   G_0(x,x_1) \, \Sigma(x_1,x_2) \, G(x_2,x') \>,
   \label{e:SDG}
\end{align}
The gap equation for $\chi(t)$ is:
\begin{equation}
   \chi(t)
   =
   \mu^2 + \frac{\lambda}{2} \, G(x,x)/i \>.
   \label{e:chi}
\end{equation}
Here $D_0^{-1}(x,x')$ and $G_0^{-1}(x,x')$ are given by:
\begin{align}
   D_0^{-1}(x,x')
   &=
   - \frac{1}{\lambda} \, \delta_{\C}(x,x')  \>,
   \\
   G_0^{-1}(x,x')
   &=
   [ \, \Box + \chi(t) \, ] \,  \delta_{\C}(x,x')  \>.
\end{align}
Thus $D_0(x,x') = - \lambda \, \delta_{\C}(x,x')$.  The exact
equations for the polarization~$\Pi$ and the self energy~$\Sigma$
are
\begin{alignat}{2}
   \Pi(x,x')
   &=
   \frac{i}{2} \, \int \rd x_1 \rd x_2 \,
   G(x,x_1) \, \Gamma(x_1,x',x_2) \, G(x_2,x) \>,
   \\
   \Sigma(x,x')
   &=
   i \, \int \rd x_1 \rd x_2 \,
   G(x,x_1) \, \Gamma(x_1,x',x_2) \, D(x_2,x) \>.
   \label{e:pisigamex}
\end{alignat}
In the BVA we make the further approximation that the \emph{exact}
one-particle irreducible $\chi \phi \phi$ vertex function $\Gamma$
is replaced by its lowest order value.
\begin{equation}
   \Gamma(x_1,x',x_2) = \delta (x_1 - x') \delta (x_2 - x') \>.
\end{equation}
so that we have in the quantum BVA
\begin{alignat}{2}
   \Pi(x,x')
   &=
   \frac{i}{2} \,
   G(x,x') \, G(x',x) \>,
   & \qquad
   \Sigma(x,x')
   &=
   i \, G(x,x') \, D(x',x) \>.
   \label{e:pisigam}
\end{alignat}
To remove the tadpole contributions we write
\begin{equation}
   D(x,x')
   =
   D_0(x,x') + \bar{D}(x,x')
   =
   - \lambda \, \delta_{\C}(x,x') + \bar{D}(x,x')  \>,
   \label{e:barDdef}
\end{equation}
which we put into Eq.~\eqref{e:SDD}, to find an integral equation for
$\bar{D}(x,x')$:
\begin{equation}
   \bar{D}(x,x')
   =
   - \lambda^2 \, \Pi(x,x')
   + \lambda \int_{\C} \rd^2 x'' \, \Pi(x,x'') \, \bar{D}(x'',x') \>.
   \label{e:barDeq}
\end{equation}
In addition, we put
\begin{equation}
   \Sigma(x,x')
   =
   \lambda \, [ G(x,x)/i ] \, \delta_{\C}(x,x')
   +
   \bar{\Sigma}(x,x') \>,
   \label{e:barSigmadef}
\end{equation}
where
\begin{equation}
   \bar{\Sigma}(x,x')
   =
   i \, G(x,x') \, \bar{D}(x',x) \>.
   \label{e:SigmabarD}
\end{equation}
Now by multiplying Eq.~\eqref{e:SDG} by $G_0^{-1}(x,x')$ gives a
differential-integral equation for $G(x,x')$:
\begin{equation}
   [ \, \Box + \chi(t) \, ] \, G(x,x')
   =
   \delta_{\C}(x,x')
   -
   \int_{\C} \rd^2 x'' \,
   \Sigma(x,x'') \, G(x'',x') \>,
   \label{e:Gdi}
\end{equation}
Putting \eqref{e:barSigmadef} into \eqref{e:Gdi} gives:
\begin{equation}
   [ \, \Box + \bar{\chi}(t) \, ] \, G(x,x')
   =
   \delta_{\C}(x,x')
   -
   \int_{\C} \rd^2 x'' \,
   \bar{\Sigma}(x,x'') \, G(x'',x') \>,
   \label{e:Gdii}
\end{equation}
where we define $\bar{\chi}(t)$ by
\begin{equation}
   \bar{\chi}(t)
   =
   \chi(t) + \lambda \, G(x,x)/i
   =
   \mu^2 + \frac{3 \lambda}{2} \, G(x,x)/i \>,
   \label{e:barchi}
\end{equation}
which is the Hartree approximation effective mass. In summary the
full set of equations are
\begin{align}
   [ \, \Box + \bar{\chi}(t) \, ] \, G(x,x')
   &=
   \delta_{\C}(x,x')
   -
   \int_{\C} \rd^2 x'' \,
   \bar{\Sigma}(x,x'') \, G(x'',x') \>,
   \label{e:Gdiii} \\
   \bar{D}(x,x')
   &=
   - \lambda^2 \, \Pi(x,x')
   + \lambda \int_{\C} \rd^2 x'' \, \Pi(x,x'') \, \bar{D}(x'',x') \>.
   \label{e:barDeqi} \\
   \bar{\Sigma}(x,x')
   &=
   i \, G(x,x') \, \bar{D}(x',x) \>.
   \label{e:SigmabarDi} \\
   \Pi(x,x')
   &=
   \frac{i}{2} \,
   G(x,x') \, G(x',x) \>,
   \label{e:Pi} \\
   \bar{\chi}(t)
   &=
   \mu^2 + \frac{3 \lambda}{2} \, G(x,x)/i \>.
   \label{e:barchii}
\end{align}
Note the factor of $3/2$ in Eq.~\eqref{e:barchii}.  We can now
\emph{redefine} $G_0(x,x')$ using $\bar{\chi}(t)$ rather
than $\chi(t)$.  We call this propagator $\bar{G}_0(x,x')$.  So let
\begin{equation}
   \bar{G}_0^{-1}(x,x')
   =
   [ \, \Box + \bar{\chi}(t) \, ] \,  \delta_{\C}(x,x')  \>.
   \label{e:barG}
\end{equation}
Then Eq.~\eqref{e:Gdiii} can be written as an integral equation for
$G(x,x')$ using $\bar{G}_0(x,x')$ and $\bar{\Sigma}(x,x')$:
\begin{equation}
   G(x,x')
   =
   \bar{G}_0(x,x')
   -
   \int_{\C} \rd^2 x_1 \, \int_{\C} \rd^2 x_2 \,
   \bar{G}_0(x,x_1) \,
   \bar{\Sigma}(x_1,x_2) \, G(x_2,x') \>,
   \label{e:Gdiiii}
\end{equation}
which is completely equivalent to the original equations, except that
now we have explicitly removed the delta-function term in $D(x,x')$
and included it in the definition of $\bar{\chi}(t)$.

%
%
\subsection{Two-particle irreducible effective action}
\label{s:TpIEA}

As we have discussed in a previous paper, the BVA can also be
obtained by keeping the two loop graph in the effective action for
the generating functional for the 2-PI graphs\cite{ref:CJT}.
Namely, the effective action is the twice-Legendre transformed
generating functional:
\begin{equation}
   \Gamma[\Phi,\G]
   =
   S_{\text{class}}[\Phi] +
   \frac{i}{2} \mathrm{Tr} \{ \, \ln \, [ \, \G^{-1} \, ] \}
   +
   \frac{i}{2} \mathrm{Tr} \{ G^{-1}[\Phi] \, \G - 1 \}
   +
   \Gamma_2[\Phi,\G]  \>.
   \label{eq:CJT}
\end{equation}
where
\begin{equation}
   G^{-1}_{\alpha,\beta}[\Phi](x,x')
   =
   - \frac{\delta^2 S_{\text{cl}}[\Phi]}
          {\delta \Phi_{\alpha}(x) \, \delta \Phi_{\beta}(x')}
\end{equation}
For our problem $\Phi$ consists of both $\phi$ and $\chi$ and is
also a matrix in CTP space. $S_{cl}$ is the classical Lagrangian
written in terms of both $\phi$ and $\chi$.  To obtain the BVA one
just keeps the two loop graph made from three $\Phi$ propagators.
For the case $ \langle \phi \rangle \neq 0$, the $\Phi$ propagator
will be non diagonal in $\phi$-$\chi$ space.  The quantity
$\Gamma_2[\Phi,\mathcal{G}]$ has a simple graphical interpretation
in terms of all the 2-PI vacuum graphs using vertices from the
interaction term.  When $\expectz{\phi} = 0$, one obtains
\begin{multline}
   \Gamma[\chi,\G_\phi,\D]
   =
   S_{\text{class}}[\chi]
   +
   \frac{i}{2} \mathrm{Tr} \{ \, \ln \, [ \, \D^{-1} \, ] \}
   \\
   +
   \frac{i}{2} \mathrm{Tr} \{ \, \ln \, [ \, \G_\phi^{-1} \, ] \}
   +
   \frac{i}{2} \mathrm{Tr}
      \{ D^{-1} \, \D + G^{-1} \, \G_\phi - 2 \}
   +
   \Gamma_2[\G_\phi,\D] \>.
   \label{eq:GammaDDSA}
\end{multline}
where $\G \equiv \{ \G_\phi,\D \}$ and
\begin{equation}
   \Gamma_2[\G_\phi,\D]
   =
   - \frac{1}{4}
    \int_\C \rd x_1 \int_\C \rd x_2 \,
      \D(x_1,x_2) \, \G_\phi(x_1,x_2) \, \G_\phi (x_2,x_1)  \>.
\end{equation}
It is often useful to put this generating functional in the Baym form
by defining $\Sigma =\G ^{-1} - G^{-1}$ to obtain:
\begin{equation}
   \Gamma[\Phi,\G]
   =
   S_{\text{class}}[\Phi] +
   \frac{i}{2} \mathrm{Tr} \{ \, \ln \, [ \, \G^{-1} \, ] \}
   -
   \frac{i}{2} \mathrm{Tr} \{ \Sigma  \, \G  \}
   +
   \Gamma_2[\Phi,\G]  \>.
   \label{eq:Baym}
\end{equation}
Eq.~\eqref{eq:Baym} allows us to relate the graphs in the self energy
$\Sigma$ to those kept in the effective action $\Gamma_2$ by means of
\begin{equation*}
   \frac{i}{2} \ \Sigma
   \ = \
   \frac{\delta \Gamma_2}{\delta \G} \>.
\end{equation*}

%
%
\subsection{Classical limit}
\label{s:classlimit}

As we discuss in the appendices, the SD equations for quantum and
classical evolutions are similar in structure, the only difference
being that the classical evolution contains fewer vertices.  The
interaction part of the CTP Lagrangian in the original matrix
bases is given by
\begin{equation}
   L_I
   =
   \half \, [ \chi_+ \phi_+ \phi_+ - \chi_- \phi_- \phi_-]
\end{equation}
Rotating to the advanced-retarded basis by the transformation
\begin{alignat}{2}
   \chi_+
   &= \chi \, + \, \frac{\hbar}{2} \, {\hat \pi}_\chi  \>,
   & \qquad
   \chi_{-}
   &= \chi \, - \, \frac{\hbar}{2} \, \hat{\pi}_\chi \>,
   \notag \\
   \phi_+
   &= \phi \, + \, \frac{\hbar}{2} \, \hat{\pi}_\phi\>,
   & \qquad
   \phi_{-}
   &= \phi \, - \, \frac{\hbar}{2} \, \hat{\pi}_\phi\>,
\end{alignat}
the Lagrangian becomes
\begin{equation}
   L_I / \hbar
   =
   \half \,
      \biggl \{
         2 \, \chi \, \phi \, \hat{\pi}_{\phi} +  \hat{\pi}_{\chi} \,
         \biggl [ \,
            \phi^2 \, + \, \frac{\hbar^2}{4} \, \hat{\pi}_\phi^2 \,
         \biggr ]
      \biggr \}
\end{equation}
so that the bare vertices of the rotated Lagrangian are:
\begin{gather}
   \gamma_{\chi \phi {\hat \pi}_\phi }
   =
   \gamma_{\phi \chi {\hat \pi}_\phi }
   =
   \ldots \text{4 perms} \ldots
   =
   \frac{1}{6} \>,
   \qquad
   \gamma_{\phi \phi \hat{\pi}_\chi}
   =
   \gamma_{\phi \hat{\pi}_\chi \phi}
   =
   \gamma_{ \hat{\pi}_\chi \phi \phi}
   = \frac{1}{3} \>,
   \\
   \gamma_{ \hat{\pi}_\chi \hat{\pi}_\phi \hat{\pi}_\phi }
   =
   \gamma_{ \hat{\pi}_\phi \hat{\pi}_\chi \hat{\pi}_\phi }
   =
   \gamma_{ \hat{\pi}_\phi \hat{\pi}_\phi \hat{\pi}_\chi }
   =
   \frac{\hbar^2}{24} \>.
\end{gather}
In the classical limit the vertices of the last line which have an
extra factor of $\hbar^2$ are missing.  The self energy in the BVA
is given by
\begin{equation}
   \Sigma_{ab}
   =
   \gamma_{ajk} \, G^{jl} \, D^{km} \, \gamma_{lmb} \>.
\end{equation}
In the above $j,l$ are summed over $\phi$ and ${\hat \pi}_\phi$
whereas $k,m$ are summed over $\chi$ and ${\hat \pi}_\chi$.  Similarly
the vacuum polarization in the BVA is given by
\begin{equation}
   \Pi_{ab}
   = \half \, \gamma_{ajk} \, G^{jl} \, G^{km} \, \gamma_{lmb} \>,
\end{equation}
where in the above $j,l,k,m$ are summed over $\phi$ and
$\hat{\pi}_\phi$.  The inverse Green's functions are directly
related to the self energies as discussed in
Appendix~\ref{s:MSRtheory}.

%
%
\section{The Effective Action in the Single Field Formulation}
\label{s:twoPIeffaction}

In the single field formulation, we consider an action given by:
\begin{equation}
   \Gamma[\phi,G]
   =
   S_{\text{class}}[\phi] +
   \frac{i}{2} \mathrm{Tr} \{ \, \ln \, [ \, G^{-1} \, ] \}
   -
   \frac{i}{2} \mathrm{Tr} \{ \Sigma  \, G  \}
   +
   \Gamma_2[\phi,G]  \>.
   \label{eq:Baym2}
\end{equation}
where now $S_{\text{class}}$ is the action with the usual $\phi^4$
self-interactions.  The Hartree approximation keeps the
two-loop contribution to $\Gamma_2[\phi,\G]$ given by:
\begin{equation}
   \Gamma_{2}[\phi,G]_{\text{Hartree}}
   =
   \frac{3 \lambda}{8} \int \rd x  \, [ \, G(x,x) \, ]^2 \>.
\end{equation}
It is easy to go beyond Hartree by using the three-loop contribution:
\begin{equation}
   \Gamma_{2}[\phi,G]_{\text{three-loop}}
   =
   \frac{\lambda^2}{16} \iint \rd x_1 \, \rd x_2
      \, [ \, G(x_1,x_2) \, ]^4
\end{equation}
This approximation was discussed in detail by Calzetta and
Hu\cite{ref:Hu}, and recently utilized by Berges and
Cox\cite{ref:berges1} to discuss equilibration in 1+1 dimensional
quantum field theory.  The approximation, which we denote as 2-PI,
is lower order than the BVA approximation in that it can be
obtained from the BVA by keeping only the \emph{first} term in the
equation for $\bar{D}(x,x')$ as a power series in $\lambda$.  That
is, if we approximate
\begin{equation}
   \bar{D}(x,x')
   \approx
   - \lambda^2 \, \Pi(x,x')
   =
   - \lambda^2 \, G(x,x') \, G(x',x) \>,
\end{equation}
in the BVA, then
\begin{equation}
   \bar{\Sigma}(x,x')
   \approx
   \frac{\lambda^2}{2} \, [ \, G(x,x') \, ]^3  \>.
\end{equation}
So \eqref{e:Gdii} reduces to:
\begin{equation}
   [ \, \Box + \bar{\chi}(t) \, ] \, G(x,x')
   =
   \delta_{\C}(x,x')
   -
   \frac{\lambda^2}{2} \,
   \int_{\C} \rd^2 x'' \,
    [ \, G(x,x'') \, ]^3 \, G(x'',x') \>.
\end{equation}
We can rewrite this as:
\begin{multline}
   \left  \{
   \Box + \mu^2 + \frac{3 \lambda}{2} \, G_{>}(x,x)/i \,
   \right \} \, G_{>}(x,x')
   \\
   =
   -
   \frac{\lambda^2}{2} \,
   \int \rd x'' \,
   \Biggl \{
   \int_0^{t'} \rd t'' \,
      G_>(x',x'') \, G_>^3(x,x'')
      +
   \int_{t'}^t \rd t'' \,
      G_>(x'',x') \, G_>^3(x,x'')
   \\
      -
   \int_0^t \rd t'' \,
      G_>(x'',x') \, G_>^3(x'',x)
   \Biggr \} \>,
   \label{e:GgBCa}
\end{multline}
which agrees with Eqs.~(9) and (10) of Berges and
Cox\cite{ref:berges1} apart from a symmetry factor of $1/3$ on
the right hand side of Eq.~\eqref{e:GgBCa}. For the classical
case, we need to only include in $\Pi_>(x,x')$ and $ \Pi_<(x,x')$
the subset of vertices that come from the classical MSR formalism
(see appendix~\ref{s:MSRtheory}). What we find is that the results
for the 2-PI approximation are quite similar to those found for
the BVA.

%
%
\section{Results}
\label{s:results}

The present codes for evolving the SD equations use spectral numerical
methods based on Chebyshev polynomial expansions, as explained
in\cite{ref:MDC,ref:BMIM}.  Thus our calculations are not lattice
calculations in coordinate space using periodic boundary conditions,
but are carried out entirely in momentum space.  They take advantage
of the global spectral convergence character of the Chebyshev
polynomial expansion, which allow for an exact continuum limit.  The
calculation of the different modes decouple and an implementation via
the message passing interface [MPI] on a parallel computer is
straightforward.

We choose our initial parameters to match those given by Aarts,
\emph{et al.}\cite{ref:abw}.  We have taken $\lambda = 1/3$ and $\mu =
1$, which are the values which make the classical problem parameter
free.  We set $\Lambda = \pi/a = 4\pi \mu = 4\pi$.  This makes $a =
1/4$.  The physical size of the system is then fixed by the value of
$N$ from the relation, $L = N a = \pi N / \Lambda = N / 4$.

Also, to make contact with the simulations of Ref.~\cite{ref:abw}, we
choose our initial density matrix to be described by
Eq.~\eqref{e:densityz} with the mass parameter set equal to the bare
mass $\mu^2$\footnote{This choice is not optimal, since the effective
mass at $t=0$ is the Hartree value.  In classical 1+1 dimensional
field theory there is only finite mass renormalization, and using
$\mu^2$ in our initial density matrix is allowable.  However, if we
were to study the quantum version of this problem, we would have to
use the renormalized mass and not the bare mass in the initial density
matrix.}.  The value of $\beta_0$ is taken to agree with the lattice
coordinate space calculation of Ref.~\cite{ref:abw}, $1/\beta_0 = T_0
= 5.03891094$, so that for the coordinate space Monte Carlo
calculations, $\expectz{ \phi^2(x,0) } = 2.5$ exactly.  (This is for
convenience only.)

For the Hartree and BVA approximations, we take:
\begin{equation}
   f_k(0)
   =
   1 / \sqrt{ 2 \omega_k } \>,
   \qquad
   \dot{f}_k(0)
   =
   - i \omega_k \, f_k(0)
   =
   - i \, \sqrt{ \omega_k / 2 } \>,
   \label{e:fdotf}
\end{equation}
where $\omega_k = \sqrt{k^2 + \chi(0)}$, with
\begin{equation}
   \chi(0)
   =
   \mu^2
   +
   \frac{3\lambda}{2} \, \expectz{ \phi^2(0) } \>,
\end{equation}
and $I(\Lambda)$ is given by Eq.~\eqref{e:Imomentumspace}.

\begin{figure}
   \centering
   \includegraphics{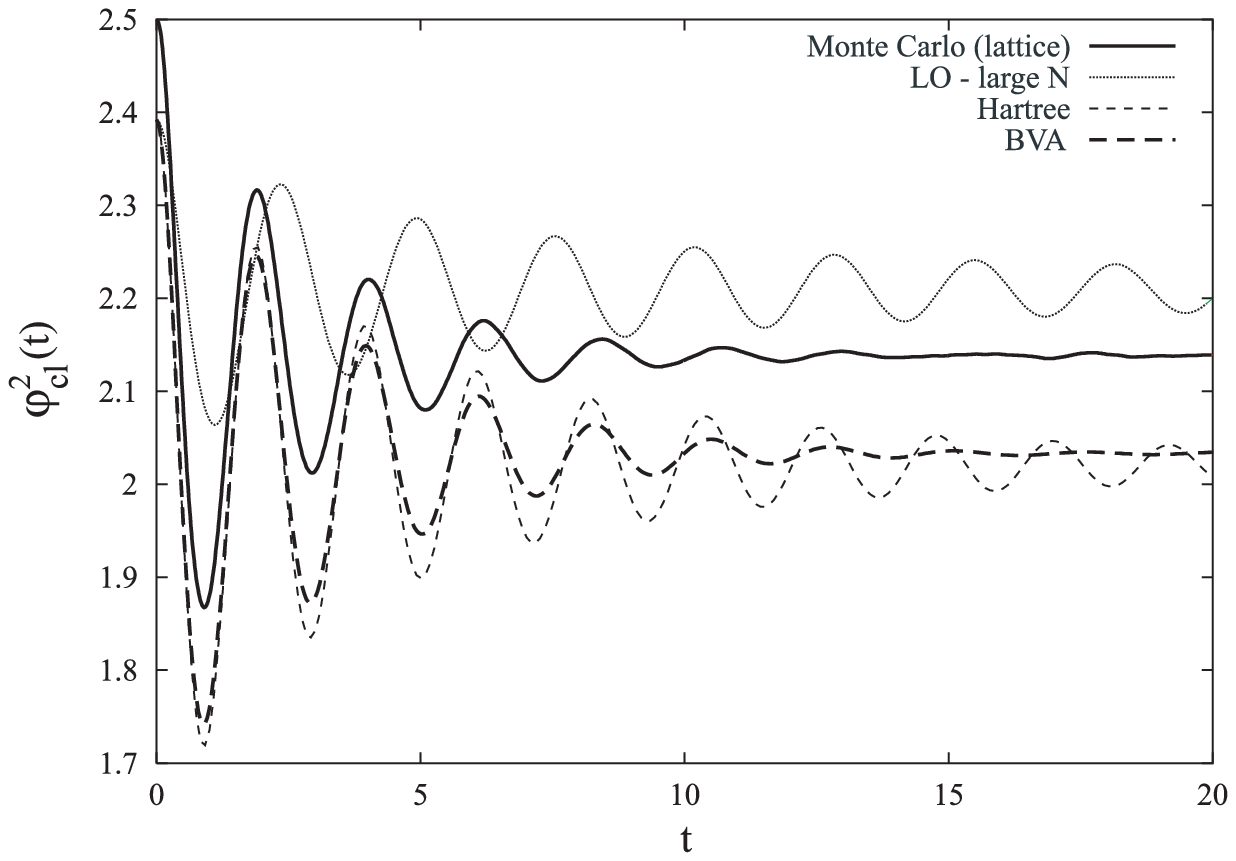}
   \caption{The lattice coordinate space Monte Carlo calculation 
   of $\phi^2_{\text{cl}}(t) \equiv \expectz{\phi^2(t)}$, compared
   with the leading order large-$N$, Hartree, and BVA approximations.}
   \label{fig:lattice}
\end{figure}

In Fig.~\ref{fig:lattice}, we show $\phi^2_{\text{cl}}(t) \equiv
\expectz{\phi^2(t)}$ as a function of time, using the coordinate space
lattice calculations of Eq.~\eqref{e:phicoordlattice} together with
the various approximations presented in this paper.  The BVA, Hartree,
large-$N$, and coordinate space Monte Carlo codes use quite modest
values of $N = 128$ or $256$, as the calculation is very insensitive to
this parameter.  The momentum space Monte Carlo code uses somewhat
larger values of $N$ to obtain high accuracy.

\begin{figure}
   \centering
   \includegraphics{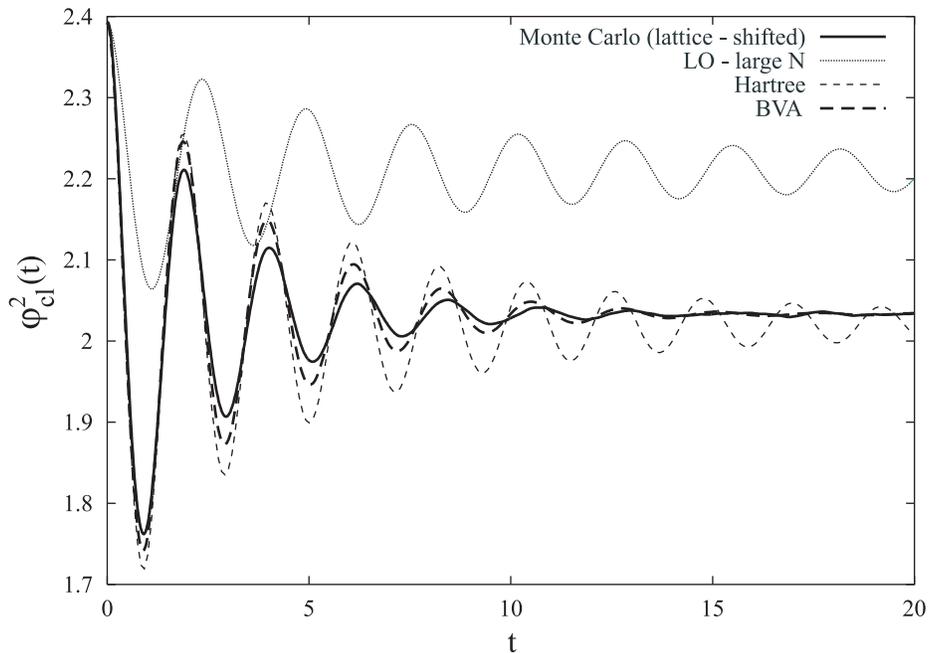}
   \caption{The \emph{shifted} lattice Monte Carlo calculation of 
   $\phi^2_{\text{cl}}(t) \equiv \expectz{\phi^2(t)}$, compared 
   with the leading order large-$N$, Hartree, and BVA approximations.  
   Here, a constant factor is subtracted from the lattice coordinate 
   space Monte Carlo results of Fig.~\ref{fig:lattice}.}
   \label{fig:shift}
\end{figure}

We see that that the initial value $\phi^2_{\text{cl}}(0)$ is
different for the lattice coordinate and momentum space approaches, as
explained in Section~\ref{s:classsols}.  For fixed values of the
parameters $\beta_0$ and $\Lambda$, the value of
$\phi^2_{\text{cl}}(0)$ is different in the two formalisms.  One can
of course force $\expectz{\phi^2(0)}$ to be the same, but then the
mismatch for finite cutoff would appear somewhere else.  Or one may
try to correct this problem by shifting the lattice result such that
it matches the continuum approximations at $t=0$.  We show the results
of such a shift in Fig.~\ref{fig:shift}.  However, such a naive
solution has no fundamental support and may be misleading.

\begin{figure}
   \centering
   \includegraphics{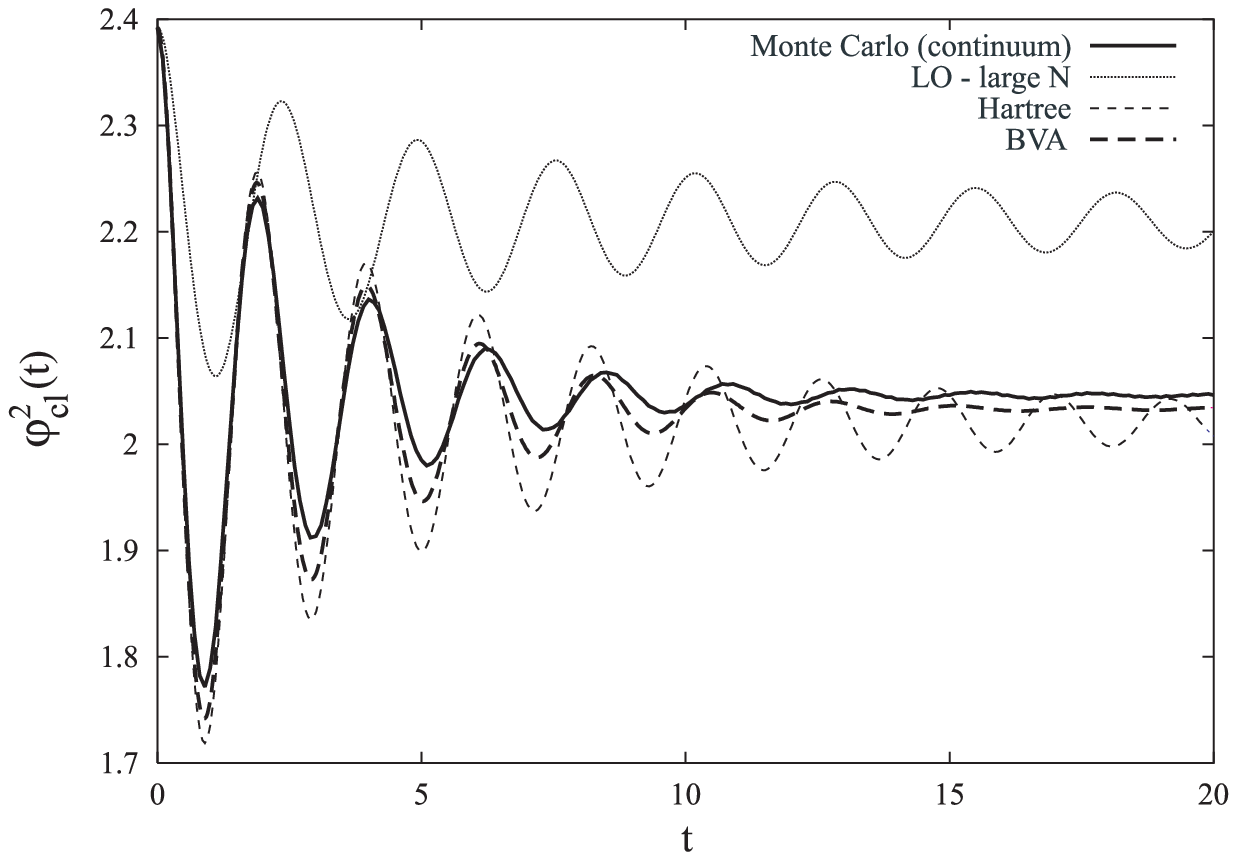}
   \caption{The lattice momentum space Monte Carlo calculation of 
   $\phi^2_{\text{cl}}(t) \equiv \expectz{\phi^2(t)}$, compared with 
   the leading order large-$N$, Hartree, and BVA approximations.}
   \label{fig:cont}
\end{figure}

\begin{figure}
   \centering
   \includegraphics{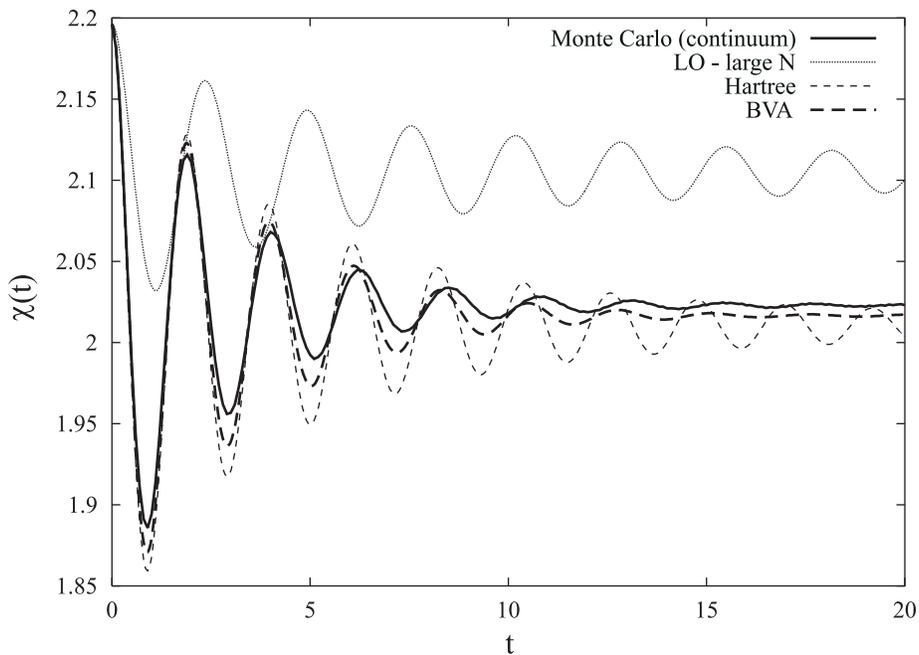}
   \caption{The lattice momentum space Monte Carlo calculation of 
   $\chi(t)$, compared with the leading order large-$N$, Hartree, 
   and BVA approximations.}
   \label{fig:chi}
\end{figure}

\begin{figure}
   \centering
   \includegraphics{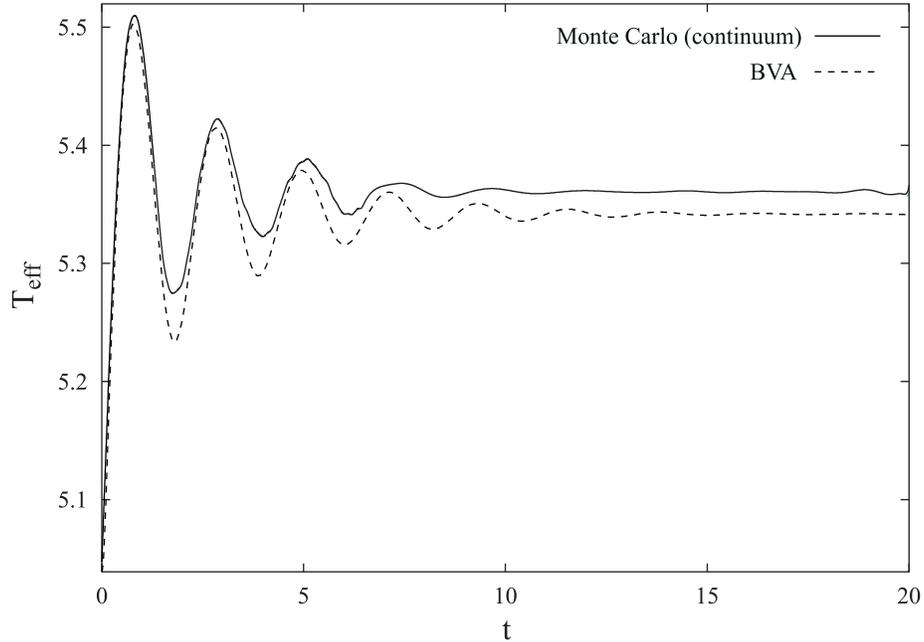}
   \caption{The effective temperature, $T_{eff}$, as defined 
   in the text, for the lattice momentum space Monte Carlo calculation
   and the BVA result.}
   \label{fig:temp}
\end{figure}

In order to compare directly our approximation methods to the exact
Monte Carlo results, we have solved the classical Monte Carlo
numerically in momentum space, as explained in
Section~\ref{s:classsols}.  In Fig.~\ref{fig:cont}, we illustrate
these lattice momentum space results for $\phi^2(t)$, and
compare them with the approximations presented here.  The BVA result
fares remarkably well and cures the oscillation problems of the
Hartree approximation at large times.  It does not display the
re-amplifications (secularity) problems manifested by the result of
Aarts \emph{et al}, where the next-to-leading order evolution is
obtained by truncating the effective action up to four-point
couplings\cite{ref:abw}.  Similar behavior is noted for the
time-evolution of $\chi(t)$, depicted in Fig.~\ref{fig:chi}, and the
\emph{effective} temperature $T_{\text{eff}} = \expect{\pi^2(t)}$,
shown in Fig.~\ref{fig:temp}.

\begin{figure}
   \centering
   \includegraphics{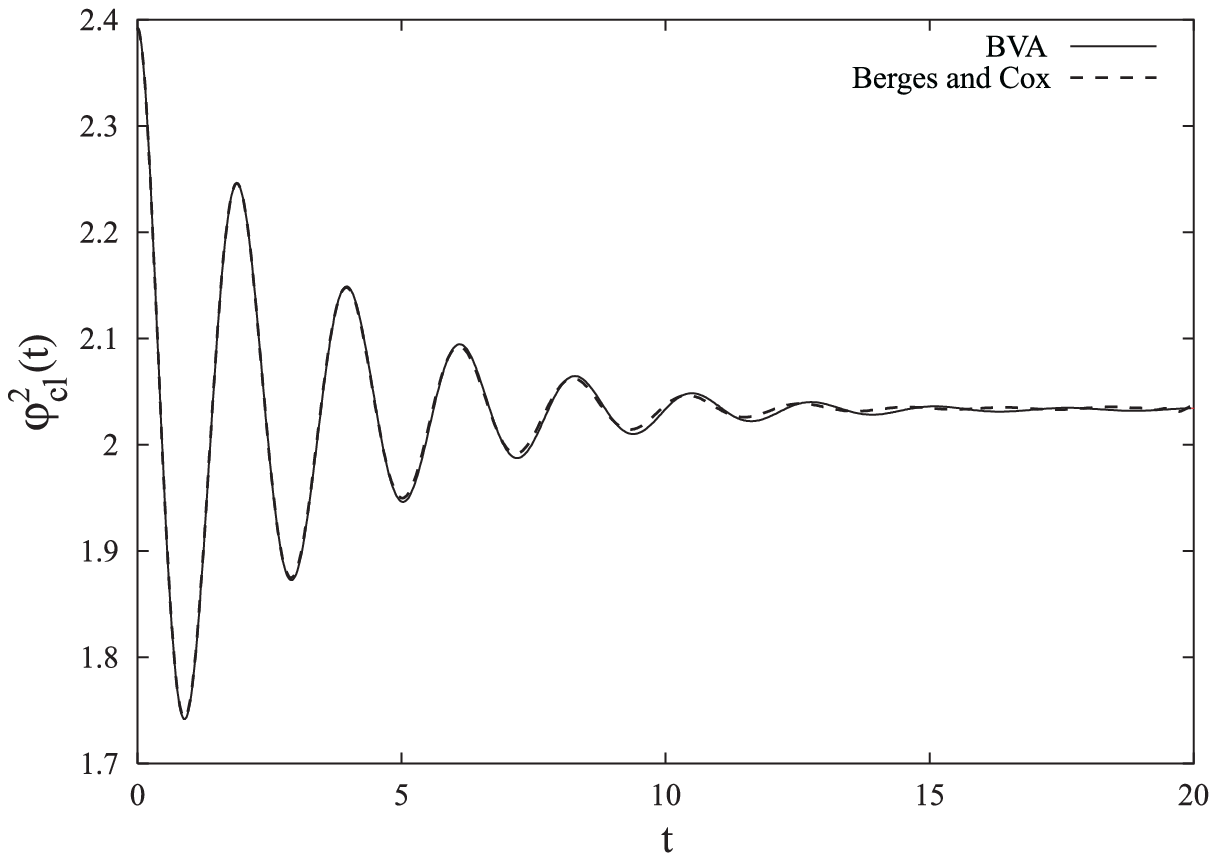}
   \caption{$\phi^2_{\text{cl}}(0) \equiv \expectz{\phi^2(t)}$ for 
   the BVA approximation and the approximation of Berges and Cox, 
   for $\mu = 1$, $\lambda = 1/3$.  Note the close agreement.}
   \label{fig:berges}
\end{figure}

In Fig.~\ref{fig:berges}, we compare the BVA result with the result of
a calculation for which we truncate the $D(x,x')$ equation at leading
order, which makes it similar to the procedure advanced by Berges and
Cox\cite{ref:berges1}.  The results are very close indeed. The reason
seems to be that the Berges approximation is expected to give
reasonable results at short times and the differences should build up
as the time-evolution progresses. However, at later times the
thermalization mechanism takes over and virtually obliterates the
differences.  Since the classical problem scales, we have only
computed results for $\mu=1$ and $\lambda=1/3$.  In the quantum case,
a second dimensionfull parameter ($\hbar$) enters, so that the problem
no longer scales, and one will have to treat the case of large
coupling constants ($\lambda=7.3$ for the linear sigma model) and high
temperatures.  Then the intermediate domain, between the regime when
the Berges' leading order approximation for $D(x,x')$ holds, and the
regime when the thermalization blurs the differences, will become more
important.

%
%

%
%
\section{Conclusions}
\label{s:conclusions}

In this paper we have compared exact numerical simulations of
classical scalar 1+1 dimensional field theory with several
Schwinger-Dyson truncation schemes.  We have focused on approximations
that do not suffer from obvious problems, such as secularity and
negative probability, by obtaining them from self-consistent
approximations to the effective action which have been shown by
Kraichnan to correspond to physically realizable systems.  In
particular, we have shown that a resummation of the $1/N$ expansion
that we studied earlier in a quantum mechanics model\cite{ref:us1},
the bare vertex approximation [BVA], cures the late time oscillation
problem of the Hartree approximation, and leads to results that
quantitatively agree with exact numerical simulations, even at late
times.  Since the classical field theory, averaged over an initial
Maxwell-Boltzmann distribution, is the high temperature limit of the
quantum field theory, averaged over an initial Bose-Einstein
distribution, our results imply that the BVA when applied to the
quantum evolution meets the minimum requirement of being valid in the
high temperature domain.

Our results give us confidence that this approximation will be useful
in future studies of quantum phase transitions in the $O(4)$ model.
The fact that in homogeneous situations this approximation leads to
thermalization, will allow us to study the rate of equilibration
versus expansion rate for an expanding plasma undergoing a phase
transition.  We will then be able to see whether some of the
interesting phenomena occurring during the early stages of a chiral
phase transition will survive the hard scatterings present in the BVA
approximation.  At the same time this study was done, a complementary
study of this approximation in the context of the quantum field theory
in 1+1 dimension was performed with results not that different from
the classical case\cite{ref:newberges}.  Again, in the quantum
simulations, the oscillations present in the Hartree approximation
were found to damp, and equilibration occurred.

%
%
\appendix
%
%

%
%
\section{CTP formalism}
\label{s:CTPtheory}

For the quantum field theory problem, the SD equations for initial
value problems can be obtained directly from the CTP path
integral\cite{ref:CTP} for the generating functional.  The
functional differential equation for the generating functional is
exactly the same as that of ordinary quantum field theory, with
the exception that the $\Theta$-functions used in defining
time-ordered products are defined on the closed-time path contour,
rather than on the real line.  This leads to a $2 \times 2 $
matrix formulation for the Green functions\cite{ref:CTP}.  It is
useful to have three different forms of the CTP formalism.  The
first is the original matrix formulation of Schwinger, the second
uses closed contour $\Theta$-functions, and the third is a rewrite
of the matrix formalism in terms of Wightman functions and
advanced and retarded propagators.

The generating functional for initial value problem Green's functions
is
\begin{equation}
   Z [J^{+}, J^{-}]
   =  e^{i W[J_{\alpha}]}
   =
   \int \prod_{\alpha}  \rd\varphi_{\alpha} \,
   \exp i \{ \,
             S[\varphi_{\alpha}] + J_{\alpha}\varphi^{\alpha} \,
          \}
   \expect{ \varphi_{1},i| \,\rho\, | \varphi_{2} , i }  \>,
\end{equation}
where $\expect{ \varphi_{1},i| \,\rho\, | \varphi_{2} , i }$ is the
density matrix defining the initial state.  We use the matrix notation
\begin{equation}
   \varphi^a
   =
   \begin{pmatrix}
      \varphi_{+} \\
      \varphi_{-}
   \end{pmatrix} \>, \qquad
   \text{for $a=1,2$.}
\end{equation}
On this matrix space there is an indefinite metric
\begin{equation}
   c_{ab} = \text{diag} \, (+1,-1) = c^{ab}
   \label{metr}
\end{equation}
so that, for example,
\begin{equation}
   J^{a} \, c_{ab} \, \varphi^{b}
   = J_{+} \, \varphi_{+} - J_{-} \, \varphi_{-} \>.
\end{equation}
~From the path integral we get the following matrix Green's
function:
\begin{equation}
   G^{ab}(t,t')
   =
   \frac{\delta^{2} W}{\delta J_{a}(t) \, \delta J_{b}(t')}
   \bigg\vert_{J =0}  \>.
\end{equation}
We can also write this using CTP $\Theta$-functions as
\begin{align}
   G(x,x')
   =
   \mathrm{Tr} \bigl \{ \,
      \mathcal{T}_{\C} [ \, \hat{\phi}(x) \hat{\phi}(x') \, ]
                     \, \}
   &=
   \Theta_{\C}(t,t') \, G_{>}(x,x') +
   \Theta_{\C}^{T}(t',t) \, G_{<}(x,x') \>,
   \notag \\
   &=
   \begin{pmatrix}
      G_{++}(x,x') & G_{+-}(x,x') \\
      G_{-+}(x,x') & G_{--}(x,x')
   \end{pmatrix} \>,
   \label{e:GSKm}
\end{align}
where the CTP $\Theta$-functions are defined by:
\begin{equation}
   \Theta_{\C}(t,t')
   =
   \begin{cases}
      1            & \text{for $t > t'$ on $\C$}, \\
      0            & \text{for $t < t'$ on $\C$},
   \end{cases}
\end{equation}
on the time contour $\C$.  In matrix form, we have:
\begin{equation}
   \Theta_{\C}(t,t')
   =
   \begin{pmatrix}
      \Theta(t-t') & 0 \\
      1 & \Theta(t'-t)
   \end{pmatrix} \>, \qquad
   \Theta_{\C}^{T}(t',t)
   =
   \begin{pmatrix}
      \Theta(t'-t) & 1 \\
      0 & \Theta(t-t')
   \end{pmatrix} \>,
\end{equation}
which leads to
\begin{align*}
   G_{++}(x,x')
   &=
   \Theta(t-t') \, G_{>}(x,x') +
   \Theta(t'-t) \, G_{<}(x,x') \>,
   \\
   G_{-+}(x,x')
   &=
   G_{>}(x,x')
   \\
   G_{+-}(x,x')
   &=
   G_{<}(x,x')
   \\
   G_{--}(x,x')
   &=
   \Theta(t'-t) \, G_{>}(x,x') +
   \Theta(t-t') \, G_{<}(x,x') \>,
\end{align*}
where
\begin{equation}
   G_{>}(x,x')/i
   =
   \Trquantum{ \hat{\phi}(x) \hat{\phi}(x') } \>, \qquad
   G_{<}(x,x')/i
   =
   \Trquantum{ \hat{\phi}(x') \hat{\phi}(x) } \>,
\end{equation}
with $\hat{\phi}(x)$ and $\hat\rho$ quantum operators.  Here,
$G_{\gtlt}(x,x')$ satisfy the symmetry relations
\begin{equation}
   \left [ \, G_{\gtlt}(x,x') / i \, \right ]^{\ast}
   =
   \left [ \, G_{\gtlt}(x',x) / i \, \right ]
   =
   \left [ \, G_{\ltgt}(x,x') / i \, \right ] \>.
   \label{e:Gsym}
\end{equation}
If we use the Schwinger-Keldysh basis then we define the
$\phi\,\phi\,\phi$ vertex as:
\begin{equation}
   \frac{g}{3} \, h_{abc} \, \phi^a \, \phi^b \, \phi^c \>,
   \qquad
   h_{abc} =  \pm 1 \>, \qquad \text{if\ \ $a=b=c=\pm 1$.}
\end{equation}
The two point function SD equation can be written as
\begin{equation}
   G^{ab} = G_0^{ab} - G_0^{ac} \, \Sigma_{cd} \, G^{db} \>.
\end{equation}
To make a direct relation with classical correlation functions,
it is convenient to go to another basis, given by
\begin{equation}
   \begin{pmatrix}
      \phi_1 \\
      \phi_2
   \end{pmatrix}
   =
   \begin{pmatrix}
      ( \phi_{+} + \phi_{-} )/2 \\
      \phi_{+} - \phi_{-}
   \end{pmatrix}
   =
   R \,
   \begin{pmatrix}
      \phi_{+} \\
      \phi_{-}
   \end{pmatrix}
\end{equation}
with
\begin{equation}
   R
   =
   \begin{pmatrix}
      1/2 & 1/2 \\
        1 & -1
   \end{pmatrix} \>.
\end{equation}
To make contact with the MSR formalism discussed below, we will choose
$\phi_1 = \phi$ and $\phi_2 = \hbar \, \hat{\pi}_{\phi}$, so that
\begin{equation}
   \phi_{\pm}
   \ = \
   \phi \ \pm \ \frac{\hbar}{2} \, \hat{\pi}_{\phi} \>.
\end{equation}
The usual matrix for the propagator
\begin{equation}
   \mathbf{G}(x-x')
   =
   \begin{pmatrix}
      G_{++}(x-x') & G_{+-}(x-x')\\
      G_{-+}(x-x') & G_{--}(x-x')
   \end{pmatrix}  \>,
\end{equation}
gets transformed into
\begin{equation}
   \mathbf{G}(x-x')
   \rightarrow
   R \, \mathbf{G}(x-x') \, R^T
   =
   \begin{pmatrix}
      iF(x-x')     & G_{\R}(x-x') \\
      G_{\A}(x-x') & 0
   \end{pmatrix}  \>,
   \label{eqpropkel}
\end{equation}
with
\begin{align}
   F(x-x')
   &=
   \half \, \expect{
      \phi(x) \, \phi(x') + \phi(x') \, \phi(x)
                   } \>,
   \notag \\
   G_{\R}(x-x')
   &=
   G_{\A}(x'-x)
   = i \, \expect{ [ \, \phi(x),\phi(x') \, ] } \,  \Theta(t-t') \>.
   \label{eqgrgaqm}
\end{align}
The last equation can be used to define the spectral function:
\begin{equation}
   \sigma(x-x')
   =
   i \, \expect{ [ \, \phi(x),\phi(x') \, ] }
   =
   G_{\R}(x-x') - G_{\A}(x-x') \>.
   \label{eqspfuqm}
\end{equation}
In the advanced-retarded basis we then get the integral equations:
\begin{align}
   F(x,x')
   &=
   F_0(x,x')
   -
   \iint \rd x'' \rd x''' \,
   \{ \,
      G_{0\,\text{A}}(x,x'') \, \Sigma_{\text{F}}(x'',x''') \,
      G_{\text{R}}(x''',x')
   \notag \\
   & \qquad +
      F_0(x,x'') \, \Sigma_{\text{R}}(x'',x''') \,
      G_{\text{R}}(x''',x')
      +
      G_{0\,\text{A}}(x,x'') \, \Sigma_{\text{A}}(x'',x''') \,
      F(x''',x')
   \}
   \notag \\
   G_{\text{A}}(x,x')
   &=
   G_{0\,\text{A}}(x,x')
   -
   \iint \rd x'' \rd x''' \,
   G_{0\,\text{A}}(x,x'') \, \Sigma_{\text{A}}(x'',x''') \,
   G_{\text{A}}(x''',x')
   \notag \\
   G_{\text{R}}(x,x')
   &=
   G_{0\,\text{R}}(x,x')
   -
   \iint \rd x'' \rd x''' \,
   G_{0\,\text{R}}(x,x'') \, \Sigma_{\text{R}}(x'',x''') \,
   G_{\text{R}}(x''',x').
\end{align}
The SD equations for our constrained $\chi$ field problem can be
obtained by considering first two propagating fields, and then taking
the composite field limit for the second field, so the bare propagator
for the second field is replaced by a delta function (see Cooper and
Haymaker\cite{ref:CGHT}).  Since the second approach is more
transparent in the MSR formalism we will follow that here. Thus we
start with the Lagrangian
\begin{equation}
   L
   =
   \half \, \Bigl \{
      (\partial_\mu \phi)^2 +
      (\partial_\mu \chi)^2 -
      m^2 \, \phi^2 -
      M^2 \, \chi^2 -
      g \, \chi \, \phi^2
   \Bigr \}
   - j \phi - S \chi.
\end{equation}
This leads to the equations of motion in the presence of external
sources:
\begin{equation}
   \bigl [ \, \Box + m^2 \, \bigr ] \, \phi(x)
   =
   - g \, \chi(x) \, \phi(x) - j(x) \>.
\end{equation}
\begin{equation}
   \bigl [ \, \Box + M^2 \, \bigr ] \, \chi(x)
   =
   - g \, \phi^2(x) / 2  - S(x)  \>.
\end{equation}
After we determine the SD equations for the two field problem, we
recover the original equations for the single $\phi^4$ field theory by
taking the composite limit where:
\begin{gather}
   \chi(x)
   =
   \mu^2 + \frac{\lambda}{2} \, \phi^2(x)
   \notag \\
   M^2
   =
   - \frac{1}{\lambda} \>, \qquad
   g = 1 \>, \qquad
   S \rightarrow \frac{\mu^2}{\lambda} \>,
   \notag \\
   D_{0\,\R}(x-y)
   \rightarrow
   - \lambda \, \delta(x-y) \>.
\end{gather}
We want to convert the Schwinger-Keldysh Lagrangian to the
advanced retarded one. Starting from:
\begin{align}
   L
   &=
   \half \,
   \Bigl \{
      (\partial_\mu \phi_+)^2 - (\partial_\mu \phi_-)^2
     + (\partial_\mu \chi_+)^2 -(\partial_\mu \chi_-)^2
     - m^2 \, (\phi_+^2 - \phi_-^2)
     - M^2 \, (\chi_+^2 - \chi_-^2)
     \notag \\
     & \qquad
     -
     g \, (\chi_+ \, \phi_+^2  - \chi_- \, \phi_-^2)
   \Bigr \}
   - j ( \phi_+ - \phi_-) - S (\chi_+ - \chi_-) \>.
\end{align}
we obtain after changing variables to $ \phi, {\hat \pi}_\phi$
\begin{align}
   - L_{\text{CTP}}/\hbar
   &=
   \hat{\pi}_\phi \, \bigl [ \, \Box + m^2 \, \bigr ] \, \phi
   +
   \hat{\pi}_\chi \, \bigl [ \, \Box + M^2 \, \bigr ] \, \chi
   +
   g \, \hat{\pi}_\phi \, \chi \, \phi
   +
   \frac{g}{2} \, \hat{\pi}_\chi \, \phi^2
   \notag \\
   & \qquad
   +
   \frac{\hbar^2}{8} \, g \, \hat{\pi}_\phi^2 \, \hat{\pi}_\chi
   -
   j_\phi \, \phi
   -
   j_\chi \, \chi
   -
   j_{\hat{\pi}_\phi} \, \hat{\pi}_{\phi}
   -
   j_{\hat{\pi}_\chi} \, \hat{\pi}_\chi  \>.
   \label{eq:lctp}
\end{align}
We will find that we get exactly the same Lagrangian in the MSR formalism
except that the term proportional to $\hbar^2$ is missing.

%
%
\section{MSR formalism}
\label{s:MSRtheory}

In the paper of Martin-Siggia-Rose [MSR]\cite{ref:MSR}, an
operator formalism was developed which allowed them to find a
generating functional for both the retarded and Wightman functions
for first order classical field equations of the type:
\begin{equation}
   \dot x(r,t) = A[x(r,t)]
\end{equation}
where $A[x(r,t)]$ is a local polynomial in the classical field
$x(r,t)$.  In the work of MSR, $A[x(r,t)]$ could contain dissipative
terms as will as prescribed noise terms.  The formalism presented in
MSR is first order in time derivatives and not apparently
covariant.  We have recently shown\cite{ref:ckr} that there is a
covariant subset of the MSR equations in terms of which all the MSR
Green's functions can be recovered.  This subset can be derived from a
second order Lagrangian formulation which can be related to the $\hbar
\rightarrow 0$ limit of the CTP formalism of Schwinger and Keldysh as
mentioned above.

%
%
\subsection{First order formalism}
\label{s:firstorderMSR}

First let us review the first order formalism.  For the statistical
classical field evolutions of $N$ interacting classical fields
$\phi_a$, $a=1,2,\ldots,N$ then $x(r,t)$ is the $2N$ component field
consisting of $\phi_a$ and the canonical momentum $\pi_a=
\dot{\phi}_a$.
\begin{equation}
   x
   =
   \begin{pmatrix}
      \phi_a \\ \pi_a
   \end{pmatrix}
\end{equation}
If, for example, we restrict ourselves to cubic interactions, then the
vector $A$ is of the form:
\begin{equation}
   A_i
   =
   c_i (r,t)
   +
   d_{ij} x_j(r,t)
   +
   e_{ijk} x_j(r,t) x_k(r,t) / 2
\end{equation}
where $i= 1,\ldots,4$.  In the MSR formalism one then introduces the
operator
\begin{equation*}
   \hat{x}(r,t) \equiv - \frac{\delta}{\delta x(r,t)}
\end{equation*}
such that the commutation rule $[x(r,t), \hat{x}(r',t)] =
\delta(r-r')$ is true.  Then if we define an operator Hamiltonian via:
\begin{equation}
   H(t) = \int \rd r' \, \hat{x}_i(r',t) \,  A_i(r',t) \>,
\end{equation}
then the equations of motion can be written in the compact form
\begin{equation}
   \dot{x}(r,t) = [x(r,t), H(t)]
\end{equation}
For the commutator to be true at all times, one needs that
$\hat{x}$ satisfies
\begin{equation}
   \frac{ \rd \hat{x}(r,t)}{\rd t}
   = [ \hat{x}(r,t), H(t) ]
\end{equation}
for consistency.  Therefore $\hat{x}(r,t)$ is a functional of $x(r,0)$
and $\hat{x}(r,0)$.  The formal solution to these equations is given
by
\begin{align}
   x(t)
   &=
   U^{-1}(t,0) \, x(0) \, U(t,0)
   \notag \\
   \hat{x}(t)
   &=
   U^{-1}(t,0) \, \hat{x}(0) \, U(t,0)
\end{align}
where
\begin{align*}
   U^{-1}(t,t_0)
   &=
   \calT \biggl \{
      \exp \biggl [ \,
         - \int_{t_0}^t H(t') \, \rd t' \,
           \biggr ]
         \biggr \} \>,
   \\
   U(t,t_0)
   &=
   \calT^{\ast} \biggl \{
      \exp \biggl [ \,
         + \int_{t_0}^t H(t') \, \rd t' \,
           \biggr ]
         \biggr \} \>.
\end{align*}
Here $\calT$ corresponds to the usual time ordered product
operation.
The meaning of the expectation value $\expect{ x(t) \,
\hat{x}(t')}$ is as follows.  Given an initial probability
function $P[x(0)]$, then
\begin{equation}
   \expect{ x(t) \, \hat{x}(t') }
   =
   \int \rd x(0) \, x[t,x(0)] \, \hat{x}[t', x(0), \hat{x}(0) ] \,
   P[x(0)]  \>.
\end{equation}
Wherever $\hat{x}(0)$ appears, it is replaced by
\begin{equation*}
   \hat{x}(0) \rightarrow - \frac{\rd}{\rd x(0)}
\end{equation*}
and it acts on everything to the right.  This definition of the
extended averaging procedure has three important properties.
\begin{enumerate}
\item The average of a product of $x$'s agrees with the conventional
definition.
\item The time dependence of $\expect{x(t) \, \hat{x}(t')}$ is
consistent with the above equations of motion.
\item The expectation value of a product of $x$ and $\hat{x}$
which has an $\hat{x}$ to the left vanishes if $P[x(0)]$ goes to
zero fast enough at large $|x|$.
\end{enumerate}
The last property is crucial for the tridiagonal form of the Green
functions and follows from the fact that
\begin{equation*}
   \int_{-\infty}^{+\infty} \frac{\rd}{\rd x} \,
   \left [ \,  x^n P(x) \, \right ] \, \rd x = 0
   \>, \qquad
   \text{if:} \qquad
   \lim_{|x| \rightarrow \infty} x^n P[x]
   = 0 \>.
\end{equation*}
Thus in particular $\expect{ \hat{x}(t') x(t) } = 0$.

The meaning of the \emph{hatted} operators is understood in terms of
the response to the system to an external source, as is discussed in
Refs.\cite{ref:MSR} and \cite{ref:ckr}.

%
%
\subsection{Second order formalism}
\label{s:secondorderMSR}

Let us first write the equations for the two coupled fields $\chi$ and
$\phi$ in first order form.  We have
\begin{alignat}{2}
   \frac{ \partial \chi}{\partial t }
   &=
   \pi_\chi  \>,
   & \qquad
   \frac{ \partial \pi_\chi}{\partial t }
   &=
   \bigl [ \,
      \nabla^2 + m^2 \,
   \bigr ] \, \phi
   +
   g \, \chi \, \phi \>,
   \notag \\
   \frac{ \partial \phi}{\partial t }
   &=
   \pi_{\phi} \>,
   & \qquad
   \frac{ \partial \pi_\phi}{\partial t }
   &=
   \bigl [ \,
      \nabla^2 + M^2 \,
   \bigr ] \, \chi
   +
   g \, \phi^2 / 2 \>.
\end{alignat}
The operator Hamiltonian which generates these equations is
\begin{multline}
   H_{\text{MSR}}
   =
   \int \rd x \,
   \biggl \{
      \hat{\phi} \, \pi_{\phi} + \hat{\chi} \, \pi_{\chi}
   \\
      -
      \hat{\pi}_{\phi} \,
      \Bigl \{
         \bigl [ \,
            \nabla^2 + m^2 \,
         \bigr ] \, \phi
         +
         g \, \chi \, \phi \,
      \Bigr \}
      -
      \hat{\pi}_{\chi} \,
      \Bigl \{
         \bigl [ \,
            \nabla^2 + M^2 \,
         \bigr ] \, \chi
         +
         g \, \phi^2 / 2\,
      \Bigr \}
   \biggr \}
\end{multline}
Following the arguments of Ref.~\cite{ref:ckr}, and using the
commutation relations, we find that the independent covariant second
order equations (adding sources) are
\begin{align}
   \bigl [ \, \Box + m^2 \, \bigr ] \, \phi + g \, \chi \, \phi
   &=  j_{\phi}  \>,
   \notag \\
   \bigl [ \, \Box + M^2 \, \bigr ] \, \chi + g \,\phi^2/2
   &=
   j_\chi \>,
   \notag \\
   \bigl [ \, \Box + m^2 \, \bigr ] \, \hat{\pi}_{\phi}
   +
   g \, \chi \, \hat{\pi}_{\phi}
   +
   g \, \phi \, \hat{\pi}_{\chi}
   &=
   j_{\hat{\pi}_{\phi}} \>,
   \notag \\
   \bigl [ \, \Box + M^2 \, \bigr ] \, \hat{\pi}_{\chi}
   +
   g \, \phi \, \hat{\pi}_{\phi}
   &=
   j_{\hat{\pi}_{\chi}}  \>.
\end{align}
Here $\hat{\pi}_\phi(x,0)$ is treated as the operator $\delta/
\delta \phi(x,0)$ when one averages over the initial probability
function in phase space.  These equations are derivable from the
Lagrangian density:
\begin{multline}
  - \calL_{\text{MSR}}
   =
   \hat{\pi}_{\phi} \,
      \bigl [ \, \Box + m^2 \, \bigr ] \, \phi
   +
   \hat{\pi}_{\chi} \,
      \bigl [ \, \Box + M^2 \, \bigr ] \, \chi
   +
   g \, \hat{\pi}_{\phi} \, \chi \, \phi
   +
   g \, \hat{\pi}_{\chi} \, \phi^2 / 2
   \\
   -
   j_{\phi} \, \phi
   -
   j_{\chi} \, \chi
   -
   j_{\hat{\pi}_{\phi}} \, \hat{\pi}_{\phi}
   -
   j_{\hat{\pi}_{\chi}} \, \hat{\pi}_{\chi}  \>.
\label{eq:lmsr}
\end{multline}
which is identical to \eqref{eq:lctp}, except that the terms
proportional to $\hbar^2$ are missing.  This is analogous to what we
showed for the quantum mechanics example in Ref.~\cite{ref:ckr}. The
extra bare vertices of the quantum field theory calculation are
obtained from the term:
\begin{equation}
   \frac{\hbar^2}{8} \, g \, \hat{\pi}_{\phi}^2 \, \hat{\pi}_{\chi} \>.
\end{equation}

%
%
\subsection{The classical SD equations}
\label{s:SDeqsMSR}

The way one derives the SD equations from the action is identical for
both classical and quantum field theory. Thus we obtain the same
structure classically, but there are fewer vertex functions. We now
derive the SD equations for a generic cubic self-interacting field
theory whether classical or quantum.  For $N$ interacting scalar
fields, we introduce the column vector $\Phi_\alpha$ composed of
$\Phi^1_i= \phi$ and $\Phi^2_i = {\hat \pi}_i$ where $i=1,2,\ldots,
N$. We also need the metric $g_{\alpha \beta}$ which is just
$\sigma^1_{\alpha \beta}$ as far as connecting the $\Phi^1$ and
$\Phi^2$ sectors.  Then we can write generically for cubic
interactions:
\begin{equation}
   \calL
   =
   \half \, \Phi^{\alpha} \, D_{0\,\alpha\beta}^{-1} \,
   \Phi^{\beta}
   +
   \frac{1}{3} \, \gamma_{\alpha\beta\rho} \,
      \Phi^{\alpha} \,  \Phi^{\beta} \, \Phi^{\rho}
   -
   J_{\alpha} \, \Phi^{\alpha} \>,
\end{equation}
where
\begin{equation}
   D^{-1}_{0\,\alpha\beta}(x-y)
   =
   g_{\alpha \beta} \,
   \bigl [ \, \Box+ m^2 \, \bigr ] \, \delta (x-y)  \>.
\end{equation}
The generating functional is formally
\begin{equation*}
   Z[J]
   =
   \expect{ \calT \{
      \exp [ \, J_{\alpha} \, \Phi^{\alpha} \, ]
                  \}  }
   = e^{W[J]}  \>,
\end{equation*}
Defining the ``classical'' field $\Phi^\alpha$ and the connected
2-point function $W_2(1,2)$ via
\begin{equation*}
   \Phi^\alpha(1)
   =
   \frac{\delta W}{\delta J_{\alpha}(1)} \>,
   \qquad
   W_2(1,2)_{\alpha \beta}
   =
   \frac{\delta \Phi_\alpha(1)}{\delta J_{\beta}(2)}  \>,
\end{equation*}
one has that in the presence of sources
\begin{align}
   D^{-1}_{0\,\alpha\beta} \, \Phi^{\beta}(1)
   +
   \gamma_{\alpha\beta\rho} \,
      \left  [ \,
         \frac{\delta \Phi^{\beta}(1)}{\delta J_{\rho}(1)}
         +
         \Phi^{\beta}(1) \, \Phi^{\rho}(1) \,
      \right ]
   &=
   J_\alpha(1)  \>,
   \notag \\
   D^{-1}_{0\,\alpha \beta} \, W^{\beta \rho}(1,2)
   +
   \gamma_{\alpha\beta\sigma } \,
      \left  [ \,
         \frac{\delta W^{\beta \sigma}(1,1)}{\delta J_\rho(2)}
         +
         \Phi^{\beta}(1) \, W^{\sigma\rho}(1,2)
         +
         \Phi^{\sigma}(1) \, W^{\beta\rho}(1,2) \,
      \right ]
   &=
   \delta_{\alpha}^{\rho} \, \delta(1-2) \>.
   \label{eq:dy1}
\end{align}
To obtain the SD equation for the inverse two point function we first
use the connection between the connected 3-point function and the 1-PI
vertex function\cite{ref:CJT} obtained by first Legendre transforming
from $J$ to $\Phi$ and using the chain rule
\begin{equation}
   \frac{\delta}{\delta J_{\alpha}(1)}
   =
   \int \rd 2 \,
   W (1,2)^{\alpha\beta} \, \frac{\delta}{\delta \Phi(2)^{\beta}} \>.
\end{equation}
We find:
\begin{align}
   \expect{ \calT \{
      \Phi_{\alpha}(1) \, \Phi_{\beta}(2) \, \Phi_{\rho}(3)
                  \} }
   &=
   \frac{\delta W_{\alpha\beta}(1,2)}{\delta J^{\rho}(3)}
   \notag \\
   &=
   \int \rd 4 \, \rd 5 \, \rd 6 \,
      W_{\alpha \alpha'}(1,4) \, W_{\beta \beta'}(2,5) \,
      W_{\rho\rho'}(3,6) \, \Gamma^{\alpha' \beta' \rho'}(4,5,6) \>,
\end{align}
where the 1-PI three point function is defined by
\begin{equation}
   \Gamma_ {\alpha \beta \rho} (1,2,3)
   =
   \frac{\delta W^{-1}_{\alpha \beta}(1,2)}{\delta \Phi^\rho(3)}~.
\end{equation}
The usual SD equation is obtained by multiplying Eq.~\eqref{eq:dy1} on
the right by $W^{-1}$ to obtain
\begin{equation}
   W^{-1}_{\alpha \beta}(1,2)
   =
   D^{-1}_{0\,\alpha \beta}(1,2)
   +
   \Sigma_{\alpha\beta}(1,2)
   +
   2 \, \gamma_{\alpha\beta\rho} \, \Phi^{\rho}(1) \, \delta(1-2) \>,
\end{equation}
with
\begin{equation}
   \Sigma_{\alpha\beta}(1,2)
   =
   \int \rd 3 \, \rd 4 \,
   \gamma_{\alpha\rho\sigma} \, W^{\rho\lambda}(1,3) \,
   W^{\sigma\nu}(1,4) \, \Gamma(3,4,2)_{\lambda\nu\beta} \>,
\end{equation}
and where
\begin{equation}
   \Gamma_{\alpha\sigma\rho}(1,2,3)
   =
   2 \, \gamma_{\alpha\sigma\rho} \, \delta(1-3) \, \delta(1-2)
   +
   \frac{\delta \Sigma_{\alpha\sigma}(1,2)}
        {\delta \Phi^{\rho}(3)} \>.
\end{equation}
However $\Sigma$ is a proper self energy and can be
related\cite{ref:baym} to the effective action $\Gamma_2[G]$ for
the 2-PI graphs of the theory via
\begin{equation}
   \Sigma_{\alpha \beta}(1,2)
   =
   2 \, \frac{\delta \Gamma_2[G]}
             {\delta G^{\alpha\beta}(1,2)} \>,
\end{equation}
so that it is just a function of the bare vertices and the full Green
functions.  Using
\begin{equation}
   \frac{\delta \Sigma_{\alpha\beta}(1,2)}
        {\delta \Phi_{\rho}(3)}
   =
   \int \rd 5 \, \rd 6 \,
   \frac{\delta \Sigma_{\alpha \beta}(1,2)}
        {\delta G^{\sigma\lambda}(5,6)} \,
   \frac{\delta G^{\sigma\lambda}(5,6)}
        {\delta \Phi_{\rho}(3)} \>,
\end{equation}
we find that
\begin{multline}
   \Gamma_{\alpha\beta\nu}(1,2,3)
   =
   2 \, \gamma_{\alpha\beta\nu} \, \delta(1-2) \, \delta(1-3)
   \\
   - \int \rd 4 \, \rd 5 \, \rd 6 \, \rd 7 \,
   \Gamma_{\alpha\rho \sigma}(1,4,5) \, G^{\rho\eta}(4,6) \,
   G^{\sigma\lambda}(5,7) \, H_{\eta\lambda;\beta\nu}(6,7;2,3) \>.
\end{multline}
The scattering kernel $H$ is defined by
\begin{equation*}
   \frac{\delta \Sigma_{\alpha \beta}(1,2)}
        {\delta G^{\sigma\lambda}(5,6)}
   =
   H_{\alpha\beta\sigma\lambda}(1,2;5,6) \>.
\end{equation*}
Self consistent approximations\cite{ref:baym} are determined by
keeping a certain class of graphs in $\Gamma_2[G]$, the sum of all
2-PI graphs made from bare vertices and full propagators.  For cubic
interactions, the BVA is obtained by keeping the graph
\begin{equation*}
   \int \rd 1 \, \rd 2 \,
   \gamma_{ijk} G^{il}(1,2) \, G^{jm} (1,2) \,
                G^{kn}(1,2) \, \gamma_{lmn}  \>,
\end{equation*}
which then leads to the self energy being the one loop diagram, and
the scattering kernel being single particle exchange.  By using the
variational definitions of $\Sigma$ and $H$ one is guaranteed an
internally consistent approximation.  In the BVA, the integral
equation for the correlation function is
\begin{equation}
   G_{ab}(1,2)
   =
   G_{0\,ab}(1,2)
   +
   \int \rd 3 \, \rd 4 \,
      G_{0\,ai}(1,3) \, \gamma_{ijk} \,
      G^{jl}(3,4) \, G^{km}(3,4) \,
     \gamma_{lmn} \, G_{nb}(4,2) \>.
   \label{e:Gabetc}
\end{equation}
Expanding \eqref{e:Gabetc}, we can write this in terms of
classical and quantum contributions, as shown in~\cite{ref:ckr}.
We identify the self energy in this approximation as
\begin{eqnarray}
   \Sigma_{in}(3,4)
   =
   \gamma_{ijk} \, G^{jl}(3,4) \, G^{km}(3,4) \, \gamma_{lmn} \>.
\end{eqnarray}

%
%
\begin{acknowledgments}
Present calculations are made possible by grants of time on the
parallel computers of the Mathematics and Computer Science Division,
Argonne National Laboratory. We thank Harvey Rose for explaining the
MSR formalism and for providing us a copy of his unpublished thesis.
JFD and BM would like to thank LANL for hospitality during part of
this work. The work of BM was supported by the U.S.\ Department of
Energy, Nuclear Physics Division, under contract No.\ W-31-109-ENG-38.
\end{acknowledgments}
%
%
%

%
%
\end{document}